\documentclass[11pt]{article}
\usepackage{mydef2col}
\usepackage{kantlipsum,widetext}
\usepackage[utf8]{inputenc}
\usepackage[T1]{fontenc}
\usepackage[autostyle=true]{csquotes}
\usepackage{regexpatch}
\usepackage{cprotect}
\usepackage{epigraph}

\usepackage[colorinlistoftodos,prependcaption,textsize=tiny]{todonotes}
\makeatletter
\xpatchcmd{\@todo}{\setkeys{todonotes}{#1}}{\setkeys{todonotes}{inline,#1}}{}{}
\makeatother

\allowdisplaybreaks[4]
\graphicspath{{./Graphs/}}
\usepackage{tikz}
\usetikzlibrary{calc,decorations.markings}
\usetikzlibrary{arrows,patterns,positioning}

\usepackage[backend=bibtex,style=authoryear,natbib=true]{biblatex}
\makeatletter
\newrobustcmd*{\parentexttrack}[1]{%
  \begingroup
  \blx@blxinit
  \blx@setsfcodes
  \blx@bibopenparen#1\blx@bibcloseparen
  \endgroup}

\AtEveryCite{%
  \let\bibcloseparen=\bibclosebracket}

\makeatother

\DeclareFieldFormat[article]{volume}{\textbf{#1}\addspace}
\renewbibmacro{volume+number+eid}{%
    \printfield{volume}%
    \setunit{\addcomma\space}%
    \printfield{number}%
    \printfield{eid}}

\newcommand{\eff}{\mathrm{eff}}

\newcommand{\Ind}{\mathbf{1}}

\newcommand{\hatE}[3]{\hat{#1}_{#2|#3}}
\newcommand{\hatEo}[1]{\hat{#1}_{0|0}}

\def\boldepsilon{\text{\boldmath$\varepsilon$}}

\def\boldxi{\text{\boldmath$\xi$}}

 \title{Marketron games:  Self-propelling stocks vs dumb money and metastable dynamics of the Good, Bad and Ugly markets}
\shorttitle{Marketron games}
\author{
\authorstyle{Igor Halperin\textsuperscript{1}\thanks{e-mail: \url{ighalp@gmail.com}, the corresponding author} \,
and Andrey Itkin\textsuperscript{2}\thanks{e-mail: \url{aitkin@nyu.edu}}
}
\newline\newline
\textsuperscript{1}
\institution{Fidelity Investments, USA.}\\
\textsuperscript{2}
\institution{Tandon School of Engineering, New York University, USA.}
}

\date{\today}

\begin{document}

\maketitle

\lettrineabstract{We present a model of price formation in an inelastic market whose dynamics are partially driven by both money flows and their impact on asset prices. The money flow to the market is viewed as an investment policy of outside investors. For the price impact effect, we use an impact function that incorporates the phenomena of market inelasticity and saturation from new money (the $dumb \; money$ effect). Due to the dependence of market investors' flows on market performance, the model implies a feedback mechanism that gives rise to nonlinear dynamics. Consequently, the market price dynamics are seen as a nonlinear diffusion of a particle (the $marketron$) in a two-dimensional space formed by the log-price $x$ and a memory variable $y$. The latter stores information about past money flows, so that the dynamics are non-Markovian in the log price $x$ alone, but Markovian in the pair $(x,y)$, bearing a strong resemblance to spiking neuron models in neuroscience. In addition to market flows, the model dynamics are partially driven by return predictors, modeled as unobservable Ornstein-Uhlenbeck processes. By using a new interpretation of predictive signals as $self$-$propulsion$ components of the price dynamics, we treat the marketron as an active particle, amenable to methods developed in the physics of active matter. We show that, depending on the choice of parameters, our model can produce a rich variety of interesting dynamic scenarios. In particular, it predicts three distinct regimes of the market, which we call the $Good$, the $Bad$, and the $Ugly$ markets. The latter regime describes a scenario of a total market collapse or, alternatively, a corporate default event, depending on whether our model is applied to the whole market or an individual stock.
}

\section{Introduction} \label{Introd}

In this paper, we present a new model of non-linear stochastic price dynamics in financial markets, where non-linearity is produced by a combined effect of money flows and financial frictions induced by these flows. These effects give rise to complex market price dynamics which are driven by flow-induced non-linear interactions alongside various \textit{pricing signals}, and a statistical noise that encodes a purely random component of a stock price change. Our model of price dynamics is designed for medium-to-long term time scales varying from days to months to years, but does not go down to the level of intraday price modeling, where microeconomic mechanisms become dominant.

The model combines three different strands in the recent literature. First, it incorporates the inelastic market hypothesis proposed in \cite{Gabaix_Koijen_2020}, and further developed by \cite{Bouchaud_2021}. These papers are focused, respectively, on the macroeconomic and microeconomic mechanisms of the phenomenon of {\it market inelasticity}. The latter amounts to the empirical evidence that each dollar invested in a stock raises the market stock price by approximately five dollars, which strongly contradicts expectations based on the classical financial theory. This observation calls for a new model where such market inelasticity would be a built-in feature.

In general, all market flows into or out of a particular stock in a unit of time (e.g. one month) can be thought of as originating from two sources: (i) trades between professional asset managers driven by portfolio rebalancing needs or alpha-search, and (ii) flows induced by the new money in the market. The latter comes from retail investors, mostly via their contributions to retirement plans such as 401(K) plans in the US, and individual investment activities.

In this paper, we build a stylized model of market dynamics that focuses on market inelasticity effects specifically driven by new money in the market, rather than all market flows. One reason for doing this is that while new money flows are arguably a minor component of market flows for any particular stock, their dependence on the market performance as a whole could be relatively easily captured, at least approximately. Indeed, retail investors bring their money to the market when it performs well, and take their money elsewhere when it does not. As such dependence of money flows on market performance produces a feedback loop due to financial friction, it gives rise to non-linear interactions in our model. Therefore, even though money market flows may constitute only a small fraction of all market flows, its simple directional effect with market moves can justify focusing only on the money flow in the market, and not on all money flows.\footnote{
More recently, Isichenko similarly suggested to refine the approach of \cite{Gabaix_Koijen_2020,Bouchaud_2021} by focusing specifically on the effects of {\it new money} in the market, as opposed to {\it all} market flows that include both new money and trades between current market participants, \cite{Isichenko}.}

The {\it second} strand in the literature related to this paper is recent work, initially developed with no connection to \cite{Gabaix_Koijen_2020,Bouchaud_2021}, on models that bring the money flows from outside market investors, and their impact on the price formation as the main focus of modeling \cite{QED,NES,MANES}. With this approach, the overall task of building a suitable model is decomposed into two simpler sub-tasks. The first one is a model of the money flows into the market, thought of as a function of the market performance. This function is conceptualized as an {\it optimal policy function for outside investors} (such as, e.g.,  contributors to their 401(K) plans, for the US market). The second sub-model is a {\it price impact} model.

While \cite{Gabaix_Koijen_2020, Bouchaud_2021,Isichenko} are focused on market mechanisms that {\it explain} market inelasticity, the paper \cite{QED} takes a "phenomenological" approach to asset price modeling, and directly {\it incorporates} such an impact into a stochastic model for the price dynamics using a linear impact function. Note that while linear impact functions were also used
in \cite{Gabaix_Koijen_2020,Bouchaud_2021,Isichenko}, this choice is mostly driven by tractability arguments, rather than being justified on the empirical grounds.
Combining a linear impact function with a policy parameterized as a polynomial in the price, the authors of \cite{QED} have obtained a non-linear Langevin model of price dynamics, also referred to as the {\it non-linear diffusion}. A potential function of this non-linear diffusion is formed by a combined effect of money flows and its friction-induced impact on the market. One of the most interesting implications of the approach of \cite{QED} is that with such non-linear dynamics,
{\it equity naturally becomes defaultable} (as it happens in real life, of course), and importantly, {\it without} invoking any additional random process.

The {\it third} strand of the literature, incorporated into our model, is the so-called 'dumb money effect' \cite{Frazzini_2008}. It amounts to the observation that once the cumulative money flow into a stock over some period of time (typically, of the order of one year) exceeds a critical value, the stock returns usually start to diminish. As was suggested in \cite{NES, MANES}, such saturation effects of money flows can be achieved if the price impact is a {\it quadratic}, rather than a linear function of the money flow. Models developed in \cite{NES, MANES} follow the same ideas as in \cite{QED}, but are focused on a different objective. Unlike the work of \cite{QED} that focused on the real-measure dynamics of stock prices and equity default risk modeling, \cite{NES, MANES} explore a stylized version of a non-linear potential function inspired by the approach of \cite{QED}, and fit parameters of this potential using the market option data. The resulting "market-implied" Langevin potential provides a way to assess such non-linear pricing models. It turns out that the approach of \cite{NES, MANES} can be viewed as a theoretically-motivated and attractive alternative to the dominant tradition of both the academic and industrial researchers of using very complex volatility models taking into account stochastic volatility or/and jump-diffusion, Levy processes etc., to fit market option data.

The new model presented in this work is also built along the lines of reasoning outlined in \cite{QED}. We decompose the problem of building the price dynamics into two sub-problems which aim to design two separate functions: (i) a money flow model as a policy function of outside investors, and (ii) a market impact function. While the same modeling paradigm is kept, both functions (suggested as a modeling choice in this work) are different from those considered in \cite{QED}, hence leading to a different final model of non-linear Langevin dynamics of the market asset prices. Most importantly, the new model introduces the price dynamics as a {\it two-dimensional} process where the first dimension $x$ carries the current market log-price, while the second dimension is reserved for a memory variable $y$ that stores information about past money flows. As it incorporates memory effects into the price dynamics, the new model is non-Markovian in the price taken alone, but Markovian in the pair $(x,y)$.

In addition to money flows and price impact as drivers of market returns, our model assumes that they are also driven by some
return predictive signals $ z_t $.
Following the influential paper \cite{GP}, it became a common practice in the finance literature to model such signals as mean-reverting Ornstein-Uhlenbeck (OU) processes, constructed by using either companies' fundamentals or historical market prices. In our model, we keep the OU framework for modeling the RP, but assume that they are {\it unobservable}. Furthermore, we give them a new interpretation, as a "self-propelling" property of the market prices, driven by performance of a company that issues the stock.

Note that in physics and biology, systems composed of particles with such self-propelling properties are known as {\it active matter}.\footnote{Active matter describes systems whose constituent elements consume energy and are thus out-of-equilibrium. Examples include flocks or herds of animals, collections of cells, and components of the cellular cytoskeleton.}. By borrowing methods developed for analysis of active matter, we provide new insights into the {\it marketron} model. Furthermore, we will highlight in due course a similarity of our non-linear model with models developed for analysis of spiking neuron activity in neuroscience, and explain that such similarities are actually quite expected, once we start to think of the market as a system with memory.

For simplicity, in this paper we explore a stylized market that has a single stock to offer to outside investors. As a proxy for this idealized setting, in real life we can consider a market index, e.g., the S\&P500, as a market with a single stock.\footnote{There are certain caveats that make modeling of a market index different from that of single stocks, see \cite{MANES}. However, these differences will not be important for the purposes of the model presented in this paper.} Extensions of this approach to a multi-asset market are left for future work.

The paper is organized as follows. Section~\ref{sect_asset_price_dynamics} provides the derivation of non-linear stochastic dynamics of a market that is driven by both money flows and their impact. In Section~\ref{Langevin-sect}, these dynamics are represented in terms of the multidimensional Langevin equation.
Accordingly, the market price dynamics are seen as a non-linear diffusion of a particle (hereafter {\it marketron}) in a two-dimensional space formed by the log-price $x$ and a memory variable $y$.
Three typical regimes - the Good, the Bad, and the Ugly markets - are introduced and analyzed using a 1D approximation of the marketron potential called {\it the D-limit}. Section~\ref{sect_stationary_solution} discusses various properties of the marketron potential, including metastability and the existence of instantons - solutions of the dynamics equations that produce transitions between metastable market states.  Section~\ref{neuron-sect} provides some alternative views on the marketron. In particular, we reveal an analogy between our model and models of spiking neurons developed in neuroscience. In addition, we build a bridge between our marketron model and models developed in physics for the analysis of complex systems commonly known as active matter. Section~\ref{sect_model_estimation} is devoted to model calibration. We calibrate a 3D version of the marketron model to time series of the S\&P500 returns using a particle filter method and local and global optimization. We show that our model calibration indeed produces a marketron potential with metastable market regimes and instanton-facilitated transitions between them. The final section concludes.

\section{Asset price dynamics with money flows and price impact}  \label{sect_asset_price_dynamics}

We consider an idealized market that consists of a single stock issued by a firm.
The market evolves in a finite time interval $t \in [0,T_p]$, where $T_p$ represents a time horizon. We assume $T_p$ to be of the order of months or years. Denote the market price of the stock at time $t$ as $S_t$. Since we have only one stock, $S_t$ in our setting is the same as the total market capitalization of the firm. We assume that the dynamics of $S_t$ are jointly driven by i) a price signal $z_t^{(1)}$, ii) an impact of money flow $u_t$ into the market, and iii) a purely stochastic contribution, for example due to "noisy traders". The money flow $u_t$ is considered a control or decision variable in our setting. In this section, it will be treated as a fixed time-dependent function.

Similarly to the approach in \cite{QED, NES, MANES}, we start with a discrete-time setting. The process of price formation is thought of as a sequence of two steps. Let $S_t$ be referred to as the price at the beginning of the time interval $[t, t+\Delta t]$, where $\Delta t$ is the time step in the discrete setting. Let $u_t \Delta t$ be the amount of money invested in the market by outside investors at the beginning of the time interval $[t, t+\Delta t]$. The instantaneous stock price right after this cash inflow event becomes $ S_t + u_t \Delta t $. After that, the new stock price grows in the remaining part of the time step $ [t, t + \Delta t] $ with a random return $ r_t $ whose predictable component depends on a constant risk-free rate $ r $, a signal $ z_t $, and a price impact function $ \mathcal{I} \left( u, S \right) $, to be specified later. These steps are represented by the following equations
\begin{align} \label{x_t_discrete_time}
S_{t+ \Delta t} &= (1 + r_t \Delta t) (S_t  +  u_t  \Delta t), \\
r_t &= r +  z_t +  \mathcal{I}(u, S) + \frac{\sigma}{\sqrt{\Delta t}} \epsilon_t, \nonumber
\end{align}
\noindent where $\epsilon_t \sim \mathcal{N}(0, 1)$ is the white noise. When writing the second line of \eqref{x_t_discrete_time}, for simplicity, we assume that the signal and the market impact linearly contribute to the return $r_t$. More sophisticated nonlinear models can also be considered while, most likely, losing analytical tractability.
The second assumption made is about the random part $\epsilon_t$ of $r_t$, which we represent by a white noise.

By substituting the second equation in \eqref{x_t_discrete_time} into the first one and taking the continuous limit
$\Delta t = dt \rightarrow 0$, we obtain the following stochastic differential equation (SDE) as a model of the price formation dynamics in the market
\beq \label{NE_GBM}
dS_{t} = S_{t} \left(r + z_t  + \frac{u_t}{S_t} + \mathcal{I} \left( u, S \right) \right) dt + S_{t} \sigma d W_{t},
\eeq
\noindent where $W_t$ is the Brownian motion. Note that this SDE coincides with the SDE for the Geometric Brownian motion (GBM) model with a signal-dependent drift $r + z_t$ in the limit $u_t \rightarrow 0$. On the other hand, without an impact function $\mathcal{I} \left(u, S\right)$ and with $u_t < 0$, \eqref{NE_GBM} coincides with the SDE obtained in the celebrated Merton optimal consumption model \cite{Merton_1971}. The latter model addresses the problem of optimal spending by retirees post-retirement.\footnote{In case $u_t < 0$, the drift in \eqref{NE_GBM} becomes mean-reverting. This, in addition, requires further specification of the stock price behavior at $S_t = 0$, see, e.g., \cite{CarrLinetsky2006} and references therein.}

Unsurprisingly, the same SDE \eqref{NE_GBM}, but this time with $u_t > 0$ (and still without an impact function $\mathcal{I}(u,S)$), arises in a pre-retirement "mirror" problem of the Merton problem: namely, the problem of optimal contributions of working individuals to their retirement plans. Clearly, while for a fixed investment or consumption policy $u_t$ the SDE is the same for both post- and pre-retirement problems, they have different objective functions and terminal conditions.

The latter problem of optimization of cash contributions to a retirement plan by a working individual was recently addressed in \cite{SCOP}.\footnote{\eqref{NE_GBM} without the impact function $\mathcal{I}(u,S)$ is identical to Eq.~(4) in \cite{SCOP}.} Unlike the setting of \cite{SCOP}, where the impact of an individual retiree's contribution on market asset prices can be safely neglected, in the present work which deals with the aggregate money flows of {\it all} outside investors, the price impact of new money plays a key role in the dynamics of the model presented below. The other difference from the setting of \cite{SCOP} is that here we engage with non-trivial pricing signals $z_t$, whose dynamics will be presented shortly.

To proceed, we start by changing the state variable that transforms the multiplicative noise in \eqref{NE_GBM} into additive noise. This is achieved by the transformation
\beq \label{x_t}
x_t = \log \left( \frac{S_t}{S_0} \right), \qquad S_0 = S_{t=0}.
\eeq
Using It{\^o}'s lemma, from \cref{x_t,NE_GBM} we obtain the SDE for the log-price $x_t$
\beq \label{NE_GBM_2}
dx_t = \left( r + z_t - \frac{\sigma^2}{2}  + \frac{u_t}{S_t(x_t)}\Ind_{S_t > 0} + \mathcal{I} \left( u, S(x_t) \right) \right) dt + \sigma dW_t,
\eeq
\noindent where the drift term is expressed in terms of the new state variable $x_t$, and $\Ind(x)$ is the indicator function.\footnote{In this case, it takes into account that at $S_t = 0$ the flow term disappears from \eqref{NE_GBM_2}.} To this end, we must specify functions $u_t$ and $\mathcal{I} \left( u, S \right)$, as well as define the dynamics of the signal $z_t$.\footnote{Equation \eqref{NE_GBM_2} suggests that the choice of functions $u_t$ and $\mathcal{I}(u,S)$ determines the model behavior in the whole range of values $0 \geq S \leq \infty$, or equivalently $-\infty \leq x \leq \infty$.} We now address these elements of our model in more detail.

\subsection{The signal model}

It is convenient to assume that the signal $z_t$ is a process of bounded variation. Therefore, we define it as follows
\beq \label{signal_z}
z_t = v f(\theta_t),
\eeq
\noindent where $v$ is a parameter that controls the amplitude of the random signal $\theta_t$, and $f(x)$ is some function of bounded variation. The signal is given by a mixture of $K > 0$ Ornstein-Uhlenbeck (OU) processes\footnote{Usually, it suffices to take $K$ between one and three.}
\beq \label{OU}
\theta_t = \sum_{i=1}^{K} w_i \theta_{it}, \qquad d\theta_{it} = k_i\left(\hat{\theta}_i - \theta_{it}\right) dt + \sigma_{iz} d Z_{it}.
\eeq
Here $w_i, \ i = 1,\ldots,K$ are weights (assumed to be time-independent), $\hat{\theta}i, k_i, \sigma_{iz}$ are mean levels, mean reversion speeds of individual components, and volatility parameters, respectively, and $Z_{it}$ are independent Brownian motions, also uncorrelated with $W_t$.

We note here that traditional financial models use mixture of OU processes in two ways:

\paragraph{Models of stochastic volatility.} It is known that return distributions implied by market stock prices or their corresponding options demonstrate fat tails that follow a power law. To address this phenomenon, the concept of rough volatility has been introduced for both stock and option markets. It was shown in \cite{GatheralJaissonRos2014} that for a wide range of assets, historical volatility time series exhibit behavior that is much rougher than that of the Brownian motion (BM), but could be represented by a fractional BM, see e.g. the references collected at \cite{RVsite}. Also, the fractional BM is an example of a self-similar process that can generate long memory and rough behavior, though not simultaneously. As shown in \cite{Harms2019}, the fractional BM can be represented as an integral over a family of OU processes. Various Markovian approximations of the rough volatility process have been constructed independently using a fixed number of OU processes, \cite{Rogers2019,ADOL_Risk,AbiJaber2019,AbiJaberElEuch2019, Itkin2024FMF,AbiJaberLi2024}, which on one hand are capable of reproducing some stylized characteristics of the market, and, on the other hand, provide additional tractability.
Another way to account for the effects of long memory is to directly introduce it into a stochastic volatility process via parameters with delay, \cite{JuliaCareo2024}.

\paragraph{Models of stochastic drift.} Alternatively to stochastic volatility, the drift of a stochastic process can also be made stochastic. For instance, in the influential Garleanu-Pederson model, a weighted average of current and future Markowitz portfolios is employed \cite{GP}. Additional desirable features of return distributions, such as fat tails and long memory, can be captured by introducing nonlinear drift terms \cite{Wada2009}. Such nonlinearities in the drift can cause processes to mean-revert rapidly from extreme values. Empirical evidence for nonlinear drift terms has been found in stochastic interest rate models by \cite{AltSahalia1996} and \cite{Bandi2012}.

\paragraph{} In this paper we follow the approach of \cite{GP} that uses a mixture of OU processes as a model of signals, but make two modifications to it: one is technical and relatively minor, while the other one has more profound consequences.

On the technical side, compared with traditional financial models such as \cite{GP}, OU processes do not enter \eqref{NE_GBM_2} directly, but rather after being transformed using a function $f(\cdot)$ in \eqref{signal_z}. This is done to achieve better tractability.

A more profound difference is a different {\it interpretation} of signals $z_t$ in our model. In traditional financial models such as \cite{GP}, signals $z_{it}$ are {\it observable} and constructed from companies' fundamentals or past price patterns (technical signals). In contrast, here we assume that signals $z_{it}$ are {\it unobservable}. In section~\ref{sect_controlled_dynamics}, we provide an interpretation of the OU dynamics in \eqref{OU} that enables employing tools developed in physics for analysis of active matter. Thus, while the {\it mathematics} of our model for signals in \eqref{OU} is identical to that used in financial models like \cite{GP}, in this paper we use a different interpretation of signals that is rooted in physics.

\subsection{The investors policy model}

As suggested by \eqref{NE_GBM_2}, the choice of investor policy $ u_t $ as a function of state variables (and possibly other variables) is the key entry to the model to be presented below. Indeed, this choice determines many properties of the model including, in particular, its behavior at a very high price level. For example, do we want the money flow to increase with the growth of $S_t$ without limits, or would it be more prudent to assume that the money supply is large but limited? Clearly, the behavior of the resulting model for very large prices (or very long times) for these two choices would be very different.

To define a proper investor policy $ u_t$, one can consider two different approaches. With the first approach, we would start with a particular model for utility function of risk-averse outside investors, and then solve a resulting Hamilton-Jacobi-Bellman (HJB) equation to find their optimal policy. The other approach is to directly specify the functional form of $ u_t $, and then fix parameters entering this expression by fitting the resulting model to data. In this paper, we
take the second approach as producing a shorter path to the final model of market dynamics, and leave exploration of utility-based methods for future work.

We choose to specify a policy $u_t$ that stays finite for the whole range of prices $ 0 \leq S_t < \infty $, to preclude any potential spurious effects of an unbounded policy on the process dynamics. Further, we want to have a policy that depends on the market performance. To smooth the policy with respect to market fluctuations, we make it adaptive not to the most recent returns, but rather to the average log-returns; or, equivalently, make it dependent on the current price $ S_t $.\footnote{In real markets, investors' decisions are driven by recent average performance, rather than just by the current price. However, in our finite-time horizon setting on the interval $ t \in [0,T_p] $, we can use the average log-return as a smoothed-out performance measure. In its turn, the latter is proportional to the gross return $ \log S_t/ S_0 $. As the last expression is a monotonic function of $S_t$, it lets us simply make the policy a function of $S_t$, or equivalently, of the current log-return $x_t$.} Therefore, we want to have a policy $ u_t $ that increases with $ S_t $ but stays finite for any $S_t$.

Based on this, let us consider the following general representation of $u_t$, viewed here as a look-ahead investment policy of all market investors combined
\beq \label{u_t_general}
u_t = c(t)  S_0 \left[ 1 + g  G(x_t) \right],
\eeq
\noindent where $c(t) \geq 0$ is a deterministic function of time, $g$ is a parameter, and $G(x)$ is a bounded monotonically increasing function of the log-price $x = x_t$.

On top of function $G(x_t)$, the policy in \eqref{u_t_general} contains two control degrees of freedom: the function $c(t)$ and parameter $g$. The idea of such parameterization is as follows. First, we want to have one part of the policy that is independent of market performance, and another part that does depend on it. The first contribution is motivated by policies that often arise as optimal solutions for a single investor based on some optimality criteria, such as utility optimization. Therefore, the first degree of control is encoded in function $c(t)$. This function may be considered a slowly varying function, and in particular, we may consider the limit when it becomes a constant parameter, $c(t) \rightarrow c$.

On the other hand, the second term $\sim g G(x_t)$ in \eqref{u_t_general} encodes a possible dependence of the policy on past and present market performance. In the limit $g \rightarrow 0$, the policy $u_t$ becomes independent of market performance. In this limit, we recover the policy used in \cite{SCOP} to model an individual retail investor, namely a retiree planner, if $c(t) = c_0 e^{\xi t}$.

When we keep $g > 0$, the policy depends on market performance in a way prescribed by function $G(x_t)$. As any value $g > 0$ "couples" the policy $u_t$ and market performance (as measured by $S_t$ or $x_t$), in what follows we use the terminology accepted in physics and refer to constant $g$ in \eqref{u_t_general} as a {\it coupling constant} parameter.

As a particular choice of function $G(x_t)$, in this paper we focus on the following parametric specification
\begin{equation} \label{W_fun}
G(x_t) = - \frac{1}{e^{x_t}  + \varepsilon g }, \qquad g \geq 0, \quad 0 < \varepsilon \leq 1,
\end{equation}
\noindent hence
\beq \label{u_t}
u_t = c(t) S_0 \left( 1 -  \frac{g }{e^{x_t} + \varepsilon g  } \right),
\eeq
\noindent where $g, \varepsilon$ are two parameters. Thus, the total set of control parameters for the investor agent is given by the tuple $(c(t), g, \varepsilon)$. Parameters $g$ and $\varepsilon$ in \cref{u_t,W_fun} are introduced to serve two different objectives. While the coupling constant parameter $g$ "couples" the policy and the price, the second parameter $\varepsilon$ serves as a {\it regularization parameter}: it controls the behavior of the policy at small $S_t$. In particular, in the limit $S_t \ll \varepsilon g S_0, \ t > 0$, we obtain $u_t \approx c(t) S_0 (1 - 1/\varepsilon) < 0$ since $0 < \varepsilon \leq 1$. Therefore, in \eqref{u_t}, the money flow into the stock can be both positive and negative. Negative money flows $u_t$ can be interpreted as retail investors selling their stocks back to professional asset managers when the market is in a downturn, and taking their savings elsewhere, e.g., to bonds or real estate investments.

It can be easily checked that if we want $u_0 \geq 0$, this imposes an additional constraint on the possible values of $g$:  $0 \leq g \leq 1$. In what follows, we will assume this constraint on the coupling constant $ g $.

The importance of allowing finite (positive or negative) money flows in the limit of small $S_t$ can again be seen from the original \eqref{NE_GBM_2}, as it controls the overall behavior of the theory in this limit. The earlier work in \cite{QED} used parameterization that differs from \cref{u_t_general,W_fun}, instead assuming that the money flow $u_t$ should approach zero at $S_t \rightarrow 0$. On the contrary, \eqref{u_t} produces a finite and negative money flow in this limit.

Moreover, according to \eqref{u_t}, the money flow $u_t$ is non-negative only if $S_t$ exceeds the value $\hat{S} = (1- \varepsilon) g S_0$, i.e., $u_t \geq 0 \, \Longleftrightarrow \, S_t \geq \hat{S}$. This suggests an interpretation of parameter $\hat{S}_t$ as a "threshold" stock price: if $S_t$ exceeds the projected price $\hat{S}$, the money flow to the stock will be positive. Otherwise, the money flow becomes negative.

 We note here that the policy \eqref{u_t} is a {\it deterministic} policy of the future values of $x_t$, as long as the parameter $g$ is non-zero. To account for the randomness or bounded rationality of an aggregate agent representing all outside investors, we could also make the policy {\it stochastic} by adding a random component to \eqref{u_t}. For example, one choice would be to add to the right-hand side (RHS) of \eqref{u_t} the term $\sigma_u S_t d Z_t$, with a volatility parameter $\sigma_u$ and an independent Brownian motion $Z_t$. However, the resulting randomness in the money flow law $u_t$ could easily be eliminated by substituting such a policy into \eqref{NE_GBM_2} and redefining the Brownian motion $W_t$ to absorb $Z_t$. While extensions of our basic model with more complex stochastic policy specifications are certainly feasible and interesting, we leave such extensions to future work and focus in this paper on the simple deterministic policy \eqref{u_t}.

\begin{figure}[ht]
\begin{center}
\includegraphics[
width=90mm,
height=60mm]{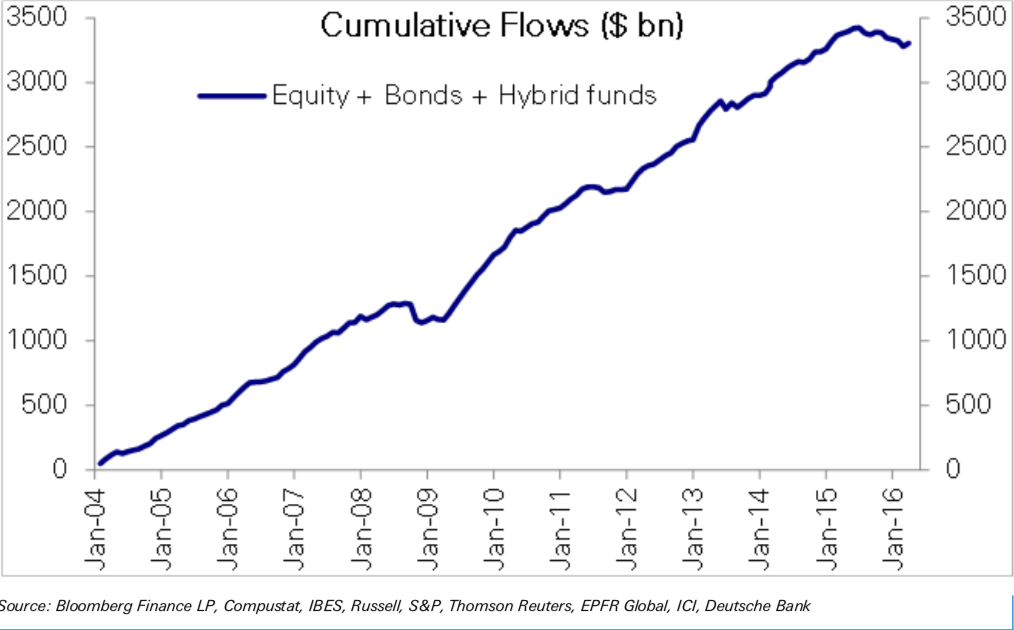}
\caption{Combined inflows into equity, bond, and hybrid funds. The annual rate is approximately constant at the level of \$325bn, \cite{DB_2016}.
}
\label{fig:fund_flows}
\end{center}
\end{figure}

The investors' policy model in \eqref{u_t} appears to be roughly in sync with data. Fig.~\ref{fig:fund_flows} demonstrates the dynamics of combined inflows into equity, bond, and hybrid funds, as per \cite{DB_2016}. It shows that on average, there was a steady inflow of around \$325bn annually into U.S. funds between 2004 and 2016, with a local drop around 2009 as a result of the economic crisis. Assuming as a rough estimate that about two-thirds of these inflows are invested in stocks, this gives rise to about \$200bn injected every year into the stock market. The main origin of such cash injection is retirement plans of U.S. workers.\footnote{Should an annual injection of \$200bn in the capital market be considered a large or negligible effect? The total market capitalization of all stocks in the S\&P500 index is about \$25.5 trillion, or \$25,500bn, so the inflows are of the order of 1\% of the total index value, which may not be a numerically insignificant effect.} Furthermore, positive correlation of market flows with market performance (achieved in our model as long as $g > 0$) is also seen on the same graph, which shows drastic changes in the flows around the economic crisis of 2008--2009, as well as during other periods of smaller-size market downturns.
\graphicspath{{./Figs/}}

\subsection{The price impact model}

A reasonable and financially justified design of the price impact function $\mathcal{I} \left( u, S \right)$ constitutes the second critical building block of the model. As will be clear from the presentation below, the functional form of the impact function, alongside the specification of the policy \eqref{u_t}, jointly determines the form of a two-dimensional potential function $V = V(x,y)$ with $x = x_t$ introduced in \eqref{x_t}, and the other variable $y = y_t$ introduced below.

When designing a model for the impact function, we want it to incorporate several related observations. First, on dimensional grounds, an {\it instantaneous} impact of a money flow at the rate $u_t$ should be proportional to the ratio $u_t/S_t$. This incorporates the market inelasticity phenomenon discussed in \cite{Gabaix_Koijen_2020, Bouchaud_2021, Isichenko}. Similarly to the approach presented in \cite{Isichenko}, we focus on money flows into the market from {\it new investors}, rather than on flows generated by trades between professional asset managers as part of their investment and risk management activities.

Second, we want the impact function to capture the "dumb money" effect, \cite{Frazzini_2008}, which amounts to diminishing stock returns once the cumulative money flow into the stock over some period of time (of the order of one year) exceeds some critical value. Such scenarios can be described as "crowded" or "saturated" scenarios.\footnote{Though stock crowding is usually addressed in terms of holding patterns for a stock across largest asset managers, intuitively we can expect that massive money flows into a particular stock by retail investors will produce holding patterns for this stock corresponding to stock crowding. Diminishing returns from crowded stocks are expected on general grounds. This occurs because during periods of market instabilities, when asset managers want to reduce their stock exposure, the stock crowdedness and resulting {\it ex-post} coordinated trading activities of asset managers produce an oversupply of the crowded stocks in the market, and hence diminishing returns. For more details, see \cite{Zlotnikov} and references therein.}

Based on this analysis, we propose the following model for the impact function $\mathcal{I} \left( u, S \right)$
\beq \label{quadr_impact}
\mathcal{I} \left( u, S \right) = y_t \frac{u_{t}}{S_{t}} \Ind_{S_t > 0},
\eeq
\noindent where $y_t$ is a new state variable that tracks the memory of past money flows into the stock.

Given that the instantaneous impact of the flow $u_t$ is proportional to the ratio $u_t/S_t$, it is natural to assume that the "memory" variable $y_t$ depends on the past values of the same ratio $u_t/S_t$. For example, one intuitive way is to define it via an exponential moving average (EMA) of some function $F = F(u_t/S_t)$
\beq  \label{exp_mov_av}
y_t =  \bar{y} - [\bar{y} - y(0)] e^{-\mu t} - \int_{0}^{t} e^{- \mu(t - t')} F\left( \frac{u_{t'}}{S_{t'}}\right) dt',
\eeq
\noindent where $\mu$ is a parameter controlling the memory depth in the market, and $0 < \bar{y} < 1$ is a threshold parameter\footnote{Since we let $c(t) \ge 0$ with no upper limit specified, and since based on \eqref{u_t} $u_t/S_t \propto c(t)$, the upper limit of $\bar{y}$ can be set to one without any loss of generality.}. Adding $y_t$ from \eqref{exp_mov_av} (or a similar one) as an additional state variable to our model aims to incorporate the "dumb money" effect of \cite{Frazzini_2008}.

Once $y_t$ falls below zero, the price impact of new money flow into the stock turns negative while $u_t > 0$. This suggests that the price dynamics in our model should be defined in terms of two variables $(x_t, y_t)$, rather than using only one variable $x_t$.
Also, in the limit $ \mu \gg 1$,
we obtain $y_t = \bar{y} - \frac{1}{\mu}F(u_t/S_t)$, which means that in this limit, \eqref{quadr_impact} reduces to a nonlinear function of $u_t/S_t$, with a small and negative coefficient in front of the nonlinear term $F\left(u_t/S_t\right)$. If we take $F(x) \propto x$, this produces a quadratic in $u_t/S_t$ term, so the price impact function obtained in this limit is similar to that previously used in \cite{QED, NES, MANES} for modeling nonlinear market dynamics.


Yet, it turns out that the choice of $F(x)$ in \eqref{exp_mov_av} is {\it not} quite arbitrary. This is because
the dynamics of $ y_t$
should be obtained as a part of the solution of a 2D Langevin equation with a certain potential $V(x,y)$. That is, the driving force for the time evolution of $y_t$ would be given by (minus) the partial derivative $(-\partial V/\partial y)$. As we will see shortly, this can be achieved by an appropriate choice of function $F(x)$.

\section{Langevin dynamics}
\label{Langevin-sect}

Plugging \eqref{quadr_impact} into \eqref{NE_GBM_2}, we put the latter into the form of a controlled Langevin equation with an external time-dependent field $z_t$ (see e.g., \cite{Coffey_book} and references therein)
\beq \label{controlled_Langevin}
dx_t = \left(z_t -  \left.
\frac{ \partial V}{ \partial x} \right|_{\substack{x = x_t, y = y_t}} \right) dt + \sigma dW_t,
\eeq
\noindent where the controlled Langevin potential is denoted as $V = V(x,y)$ (we omit control variables here for brevity), and is determined in terms of its partial derivative with respect to $x$ as follows
\beq \label{Langevin_pot_x_gradient}
\frac{\partial V}{\partial x} = - \eta - c(t) y  e^{-x} \left[ 1  + g G(x) \right] \Ind_{x \ne -\infty} = - \eta - c(t) y  e^{-x} \left( 1  - \frac{g}{ e^{x} + \varepsilon g} \right) \Ind_{x \ne -\infty}.
\eeq
Here we introduced a new parameter $\eta$ as
\beq \label{theta_param}
\eta =  r - \frac{\sigma^2}{2}.
\eeq
Note that $\eta$ can be of either sign, depending on the volatility $\sigma$ and risk-free rate $r$. Also, in what follows, we omit $\Ind_{x \ne -\infty}$, assuming $x_t > -\infty$.

Integrating \eqref{Langevin_pot_x_gradient} with respect to $x$ yields the potential $V(x,y)$ in the form
\beq  \label{Lang_pot_general}
V(x,y) = - \eta x  +  c(t) y V_{M} (x) + V(y),
\eeq
\noindent where $V(y)$ is a yet unspecified function of the second state variable $y = y_t$, and
\beq \label{U_pot_exact}
V_{M}(x) = - \int_{0}^{x} e^{-k} \left( 1 - \frac{g}{e^{k} + \varepsilon g} \right) dk =
\frac{1}{\varepsilon} \left[ (\varepsilon -1) \left(e^{-x}  -1 \right)  + \frac{1}{\kappa}  \log \frac{1 + \kappa e^{-x}}{1 + \kappa} \right], \quad \kappa = g \varepsilon.
\eeq

To make it consistent with the behavior of $y_t$ as per \eqref{exp_mov_av}, we define $V(y)$ as
\begin{equation}
V(y) = \frac{1}{2} \mu \left(y - \bar{y}\right)^2,
\end{equation}
\noindent where $\mu$ is the same parameter as in \eqref{exp_mov_av}. This produces the following 2D potential
\beq \label{Langevin_potential_exact}
V(x, y) = -  \eta x   + c(t) y V_{M}(x) + \frac{1}{2} \mu \left(y - \bar{y} \right)^2.
\eeq

Despite its appearance, the function $V_{M}(x)$ in \eqref{U_pot_exact} is non-singular in $\varepsilon$ in the limit $\varepsilon \rightarrow 0, \kappa e^{-x} \ll 1$. Indeed, expanding the square brackets in \eqref{U_pot_exact} into series in $\varepsilon$, we obtain
\beq \label{U_pot_appr}
\bar{V}_{M}(x) =   \frac{g}{2} - 1 + e^{-x} - \frac{g}{2} e^{-2x}  + O( \varepsilon ).
\eeq
This expression can be recognized as the inverted Morse (IM) potential known from quantum mechanics, \cite{Landau_QM}. Given this similarity, we will occasionally refer to $\bar{V}_M(x)$ in \eqref{U_pot_appr} as the IM potential.

In the opposite limit $x \rightarrow - \infty$, the approximation \eqref{U_pot_appr} decays as $g e^{-2x}$, while the "exact" expression \eqref{U_pot_exact} decays as $(-1/\varepsilon) e^{-x}$. This is, however, a minor issue: since the behavior of the model in the strict limit $x \rightarrow -\infty$ is determined by details of regularization, we do not try to get accurate dynamics in this region anyway, and the qualitative similarity of the model behavior in the region $x < 0$ is sufficient for our objectives. On the other hand, the approximation \eqref{U_pot_appr} is more analytically tractable than \eqref{U_pot_exact}, and thus will be our modeling choice going forward for qualitative analysis of the model behavior. However, in Section~\ref{sect_model_estimation}, when calibrating the model, we will return to the exact expression in \eqref{Langevin_potential_exact}.

Therefore, for qualitative analysis in this section, we proceed with the following 2D potential for the Langevin dynamics
\beq \label{Langevin_potential}
V(x, y) = - \eta x   + c(t) y \left[ \left(e^{-x} - 1\right) - \frac{1}{2} g \left(e^{-2x} - 1\right) \right] + \frac{1}{2} \mu \left( y - \bar{y}\right)^2.
\eeq
This expression further can be slightly modified to make the model more tractable. In particular, in the qualitative analysis in this section, we will drop the constant terms in the square brackets in \eqref{Langevin_potential}.

To complete the model, we need to define the dynamics of the memory variable $y_t$. While researchers typically treat the impact function as a deterministic function of the selling rate, \cite{HuBian2013} recognize that market liquidity is not static. They introduce a liquidity factor as a stochastic process that may represent market volume, number of market traders, and similar variables.

For markets where liquidity shows no specific trend within a given time interval, they model the market liquidity factor as a log-normal process. This approach is validated by analyzing volume data from fifty stocks in the U.S. equity market, using the Lilliefors test to check if the logarithm of the volume is a normal random variable. Their results showed that more than half of the volumes indeed follow the lognormal distribution.

The authors also consider a mean-reverting process for the liquidity factor, suggesting that liquidity tends toward a long-term stable value. This formulation of the impact function implies the existence of an upper bound on liquidity, beyond which additional selling has no influence on the stock price.




Proceeding along similar lines, in this paper we describe the memory variable $y_t$ as a stochastic process represented by the Langevin equation
\beq \label{Langevin_y}
dy_t = \left(z_t^{(2)} -  \frac{ \partial V}{ \partial y} \right) dt + \sigma_y d \tilde{W}_t.
\eeq
Here, $z_t^{(2)}$ is a random process, $\sigma_y$ is its volatility, and $\tilde{W}_t$ is the Brownian motion uncorrelated with $W_t$. We now fix $z_t^{(2)}$ to be expressed in terms of the same OU driver $\theta_t$ that appears in \eqref{signal_z}. To this end, we set $z_t^{(2)} = v h(\theta_t)$, where $h(x)$ is a function of bounded variation, to introduce a 2D vector ${\bf z}_t = \left(z_t, z_t^{(2)} \right) = v {\bf n}_t$, where ${\bf n_t} = (f(\theta_t), h(\theta_t))$. Further, by introducing the 2D Brownian motion ${\bm W}_t = \left( W_t, \tilde{W}_t \right)$ and ${\bm \sigma} = (\sigma, \sigma_y)$, two Langevin equations for $x_t$ and $y_t$ can now be combined into a vector-valued Langevin equation for the 2D state ${\bf x}_t = \left(x_t, y_t \right)$
\beq \label{vector_Langevin}
d {\bm x} = \left[ v {\bm n}_t - \nabla V({\bm x}) \right] dt  + {\bm \sigma} d {\bm W}_t.
\eeq

\begin{figure}[ht]
\begin{center}
\includegraphics[width=0.7\linewidth]{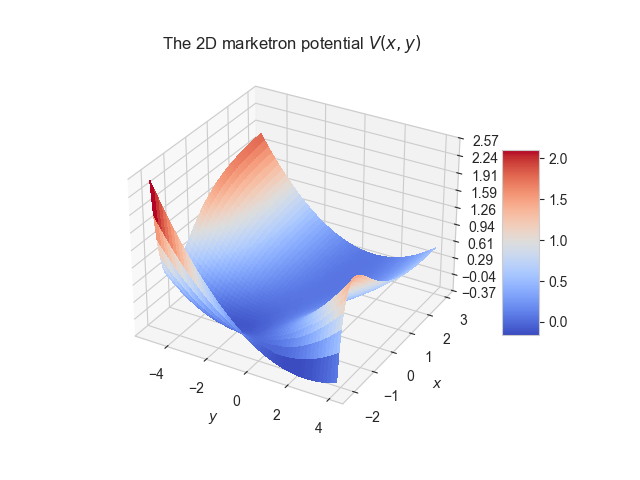}
\caption{
The marketron potential \eqref{Langevin_potential} as a function of the log-price $x$ and the memory variable $y$, for $\eta < 0$.
Parameters' values are: $c(t) = 0.13, g = 0.3, \mu = 0.1, \bar{y} = 1., \eta = -0.01$. The landscape of the marketron potential describes possible market regimes (see the main text).
}
\label{fig_Langevin_2D_pot_3D}
\end{center}
\end{figure}

\begin{figure}[ht]
\begin{center}
\includegraphics[width=175mm,height=65mm]{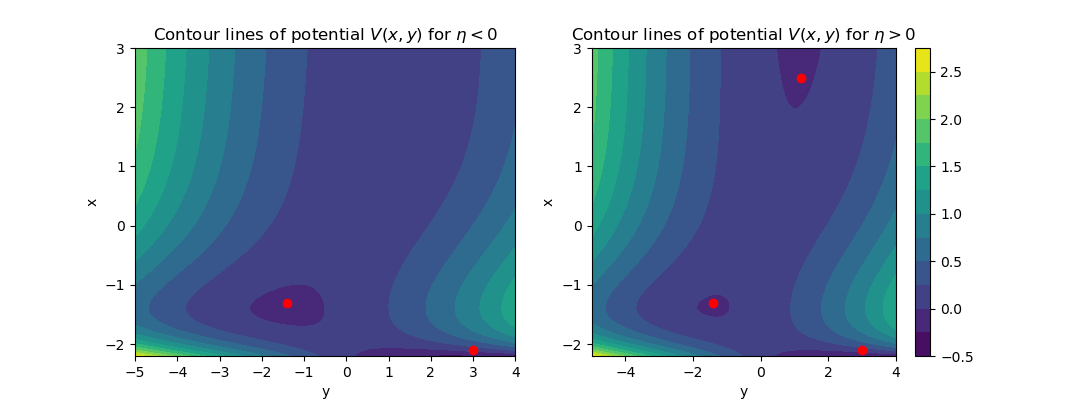}
\caption{
Contour plots of the marketron potential \eqref{Langevin_potential} for $\eta < 0$ and $\eta > 0$.
Parameters' values are: $c(t) = 0.13, g = 0.25, \mu = 0.1, \bar{y} = 1., \eta = -0.01$.
When $\eta < 0$, the particle placed in the local minimum will escape to the low right corner by barrier crossing. For $\eta > 0$, there is also an option to escape to the region $x \rightarrow \infty$, shown as an additional red dot on the top.
}
\label{fig_Potential_contour_plot}
\end{center}
\end{figure}

In what follows, the potential \eqref{Langevin_potential} will be referred to as the {\it marketron potential}. It describes a Morse-like nonlinear oscillator with the IM potential $V_M(x)$ defined in \eqref{U_pot_appr}, and an additional quadratic term, which is coupled to a harmonic oscillator. Moreover, the two oscillators are coupled {\it nonlinearly}, so that the harmonic $y$-oscillator controls the overall shape of the Morse-like $x$-oscillator. As we will see shortly, this property of the marketron potential has very interesting implications. Depending on the model parameters, the marketron potential can take various forms, such as those shown in Fig.\ref{fig_Langevin_2D_pot_3D}. Also, contour plots of the marketron potential for two cases $\eta < 0$ and $\eta \geq 0$ are shown in Fig.\ref{fig_Potential_contour_plot}.

Given the marketron potential \eqref{Langevin_potential}, the Langevin equation for $y_t$ can be made more transparent
\beq \label{Lang_y_2}
dy_t =  \left[ z_t^{(2)} + \mu (\bar{y} - y_i) -  c(t) V_M(x_t) \right] dt + \sigma_y d \tilde{W}_t.
\eeq
This is simply an OU process with a stochastic mean reversion level that also depends on $x_t$. Integrating this SDE while omitting the signal $z_t^{(2)}$ for simplicity and taking into account \eqref{u_t}, we obtain
\begin{align} \label{Lang_y_sol}
y_t &= y(0)  \bar{y} - [\bar{y} - y(0)] e^{-\mu t} - \int{0}^{t} e^{- \mu(t - s)} c(s) V_M(x_s) ds + \sigma_y  e^{- \mu t} \tilde{W}_t.
\end{align}
This expression for $y_t$ in the zero-noise limit $\sigma_y \rightarrow 0$ coincides with that in \eqref{exp_mov_av} if we choose $F(x_t) = c(t) V_M(x_t)$.

The Langevin dynamics \eqref{Lang_y_2} of the memory variable can be considerably simplified in the limit of zero noise ($\sigma \to 0$), zero signal ($z_t^{(2)} \to 0$), and short memory $\mu \gg 1$. In this limit (let us call it the {\it D-limit}), the solution \eqref{Lang_y_sol} can be approximated by a simpler deterministic law
\beq \label{y_t_large_mu_0}
y_t = \bar{y} + \frac{c(t)}{\mu} \bar{V}_M(x_t).
\eeq
As will be discussed in more detail in the next section, working in the D-limit considerably simplifies the model and reduces it to one-dimensional dynamics. However, before moving on to this discussion, we would like to make a few comments on the general properties of the 2D marketron potential in \eqref{Langevin_potential}.

\begin{figure}[ht]
\begin{center}
\includegraphics[width=95mm,height=65mm]{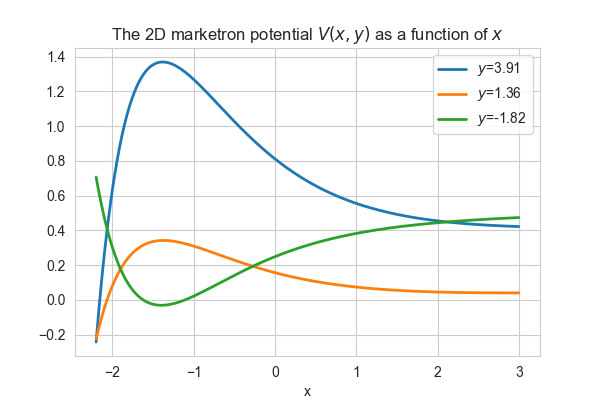}
\caption{
The marketron potential \eqref{Langevin_potential} as a function of $x$ with $\eta < 0$ for fixed values of $y$.
Parameter values are: $c(t) = 0.13, g = 0.25, \mu = 0.1, \bar{y} = 1., \eta = -0.01$. When $y > 0$, the potential enables escape either to $x \rightarrow -\infty$ or to $x \rightarrow \infty$ through a barrier whose height grows with $y$. For negative values $y < 0$, the potential becomes confining in both directions. This illustrates the 'dumb money' effect: $y < 0$ is obtained when too much money is invested in the stock. This prevents the stock from a collapse (escape to $x \rightarrow -\infty$) but also simultaneously prevents the stock from growth.}
\label{fig_Langevin_potential_vs_x}
\end{center}
\end{figure}

First, consider the limit $x \to \infty$ where the behavior of \eqref{controlled_Langevin} depends on the sign of $z_t + \eta$. If $z_t + \eta > 0$, signals are strong enough to push the price to $x \rightarrow \infty$, where the marketron potential in \eqref{Langevin_potential} becomes linearly decaying. In this limit, the potential becomes that of a free Brownian particle $V_0(x) = - \eta x$. This is, of course, as expected, because our policy is bounded at $S \rightarrow \infty$. However, in contrast to the free Brownian particle that has a linearly decreasing potential everywhere, in our case there exists a barrier that separates this asymptotic regime of a linear potential from the pre-asymptotic nonlinear regime. On the other hand, if $z_t + \eta < 0$, i.e., $z_t < -\eta$, the dynamics in \eqref{controlled_Langevin} pushes the price to $x \rightarrow -\infty$, so the marketron potential becomes a linearly increasing function of $x$ at $x \rightarrow \infty$. Therefore, in this regime the model does not admit an escape to positive infinity, as illustrated in Fig.~\ref{fig_Langevin_potential_vs_x}.

In the opposite limit $x \to -\infty$, the behavior of the model does not depend on the value of $\eta$ but instead shows a critical dependence on the value of $y$ in the potential function \eqref{Langevin_potential}. Two cases, $y > 0$ and $y < 0$, can be identified and investigated.

If $y > 0$ and $x \rightarrow -\infty$, this behavior depends on whether $g = 0$ or $g > 0$. In the strict limit $g = 0$, the potential grows to infinity, preventing escape to the negative infinity $ x \rightarrow - \infty$.
The same potential was obtained in \cite{SCOP} for a Langevin equation describing a single retirement planner agent.

However, if even an arbitrarily small value $g > 0$ is allowed, the behavior of the potential $V(x,y)$ at $x \rightarrow -\infty$ and with $y$ fixed is very different, as now it becomes unbounded and tends to negative infinity. In addition, in this regime there appears a barrier in the potential function which, provided the initial value $x_0$ is large enough, prevents a rapid escape to $x \rightarrow -\infty$ that would be observed in the absence of a barrier. As will be explicitly shown below, the height of this potential barrier depends on the model parameters as well as the value of $y_t$. In particular, parameters may be chosen in such a way that the barrier becomes extremely tall, which would make the escape to $x \to -\infty$ an arbitrarily low-probability event.\footnote{The drastic change in the model behavior depending on whether $g = 0$ or $g > 0$, even though being arbitrarily small, suggests a nonanalytic behavior of the model as a function of the parameter $g$. This will become transparent in what follows.}

Furthermore, for negative values of $y$, the potential $V(x,y)$ with $g > 0$ goes to positive infinity as $x \to -\infty$, providing confining dynamics for both limits $x \to \pm \infty$. With this scenario, $V(x,y)$ is either a single-well or double-well potential whose $x$-component is given by the sum of a Morse potential and a linear potential, see Fig.~\ref{fig_Potential_contour_plot}. On the other hand, for $g = 0$ and $y < 0$, the potential becomes unbounded at $x \to -\infty$, which again indicates non-analyticity of our model in the coupling constant $g$.

\subsection{Dynamics in the D-limit: the Good, Bad and Ugly markets} \label{sect_The_Good_Bad_Ugly}

\begin{figure}[!htbp]
\begin{center}
\includegraphics[width=90mm,height=60mm]{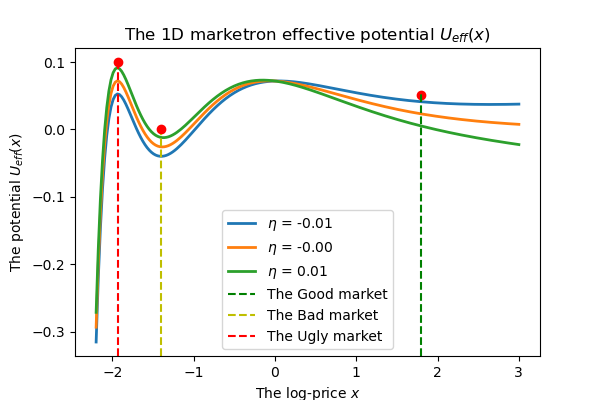}
\caption{
The effective 1D marketron potential \eqref{pot_eff} with various values of $\eta$. Larger values of $\eta$ mimic the effect of an increasing active signal $z_t$ on the overall dynamics via $\bar{\eta}$. If $\eta$ is negative, the particle can escape to the negative infinity via instantons, see \cref{sect_stationary_solution}. Red dots mark various market regimes. The Good and Bad markets are identified as regimes of fluctuations, respectively around the right-most and the mid-point red dots, positioned at the two local minima of the potential. The Ugly market corresponds to the left-most red dot and describes a market collapse scenario.
Marketron parameters' values are: $c(t) = 0.13, g = 0.25, \mu = 0.1, \bar{y} = 1., \eta = \pm0.01$.}
\label{fig_Langevin_pot_1D_eff}
\end{center}
\end{figure}

The 2D marketron dynamics (the dynamics described by a Langevin equation with the 2D marketron potential) can be considerably simplified in the D-limit defined right before \eqref{y_t_large_mu_0}. Indeed, substitution of $y_t$ from \eqref{y_t_large_mu_0} back into \eqref{Lang_pot_general} yields the marketron effective one-dimensional potential $U_\eff(x)$\footnote{In this section, to ease the notation, by $V_M(x)$ we understand its approximation $\bar{V}_M(x)$ given by \eqref{U_pot_appr}.}
\beq \label{pot_eff}
U_\eff(x) = -  \eta x + c(t) \bar{y} V_M(x) - \frac{c(t)^2}{2 \mu}  V_M^2(x).
\eeq
The effective 1D dynamics arising in this limit is now described by the Langevin equation
\beq
\label{Langevin_1D}
dx_t = \left( z_t - \frac{\partial U_\eff(x)}{\partial x} \right) dt + \sigma d W_t.
\eeq
Treating the active force $z_t$ as an effective shift of the parameter $\eta$ to $\bar{\eta} = \eta + z_t$ in \eqref{pot_eff} helps to visualize the impact of the active force $z_t$ on the resulting price dynamics. This is illustrated in Fig.~\ref{fig_Langevin_pot_1D_eff}, which shows variations in shapes of the resulting potential as the parameter $\eta$ varies.

The most interesting implication of the potential shapes, such as those shown in Fig.~\ref{fig_Langevin_pot_1D_eff}, is that they predict {\it three market regimes} (which we call the Good, the Bad, and the Ugly ones) marked by red dots on the graph. The first two dots are located at the local minima of the potential, while the leftmost dot (the Ugly) is positioned at the maximum point.

All these regimes correspond to different states of the market, thought of as regimes of fluctuations of a test particle in the effective potential \eqref{pot_eff}. The Good market scenario corresponds to diffusion around the right minimum when $x > 0$. The Bad market corresponds to the second local minimum corresponding to $x < 0$. A barrier between these minima has a peak around $x = 0$. The probability of transitions between these two regimes is controlled by the height of the barrier.

Finally, the Ugly market scenario corresponds to the leftward negative jump of the particle from the middle well (corresponding to the Bad market) over its left barrier, thus producing an escape to negative infinity in the $x$-space, or equivalently, to the zero level in the original price space. Such an escape describes a regime (or rather an event) of market collapse or equity default, depending on whether we use our model for the market as a whole or for a single stock. The next section describes dynamic mechanisms that implement such transitions between different market regimes.

\section{Properties of the marketron potential: instantons and metastability} \label{sect_stationary_solution}



To summarize so far, our model amounts to a non-linear diffusion of a particle (the marketron) in a non-linear two-dimensional potential that, depending on parameters, can have a few local minima that we
called the Good, Bad and Ugly markets. In this section, we discuss the
dynamics of transitions between these states, which are made possible by
the so-called {\it instantons} - special solutions arising in nonlinear models in statistical and quantum physics. After presenting a short overview of instantons oriented to non-physicists, we explain how they give rise to metastable dynamics of the Good, Bad and Ugly states in our model.


\subsection{Instantons}

Instantons are special transitions in non-linear statistical or quantum systems that occur between different local minima of a system's energy function, driven by either thermal or quantum fluctuations. Consider a system with a potential function like the one shown in Fig.~\ref{fig_Langevin_pot_1D_eff}. In classical mechanics, a particle placed near the Bad market equilibrium point would remain there indefinitely unless given enough external energy to overcome the barriers separating it from either the Ugly market (to the left) or the Good market (to the right).

However, in statistical and quantum physics, these transitions can occur spontaneously without external energy input, purely through internal energy fluctuations arising from thermal or quantum noise. These spontaneous barrier-crossing events are called instantons. In classical systems, instantons represent trajectories where thermal fluctuations provide enough energy for a particle to traverse a potential barrier. In quantum systems, it is quantum rather than thermal fluctuations that occasionally become strong enough to enable the particle to tunnel through the barrier.



Instantons become a good approximation to such barrier hopping transitions when the noise is small (for statistical physics case), or equivalently in a small coupling regime with a  barrier that is high relative to a typical energy of quantum oscillations, in the case of quantum tunneling. Nevertheless, even though instantons arise in a small noise (or small coupling) regime, they are essentially {\it non-perturbative} phenomena: their probabilities are non-analytical in a coupling constant, and thus cannot be found using methods of perturbation theory in a weak coupling regime \cite{Coleman_book, Zinn-Justin-QFT}.


As explained in \cite{LI}, for Langevin dynamics, instantons are obtained as solutions of classical Euler-Lagrange equations of motion that provide the leading contribution in the weak noise limit to the path integral over trajectories. For the Langevin dynamics with only
two fields $ x_t$ and $ y_t$, the instanton equations look similar to the original Langevin equations without diffusion terms and, most importantly, with the {\it flipped sign} of the gradient of the potential, to yield
\beq \label{instanton_eq_gen}
\dot{x} = \frac{\partial V}{\partial x}, \qquad \dot{y} = \frac{\partial V}{\partial y}.
\eeq
Due to the "wrong" positive sign of the potential gradient in \eqref{instanton_eq_gen}, we can also interpret these equations as a zero-noise limit of the Langevin equation, but taken in the {\it reverse time}. For a crash review of the path integral formulation
in our problem with an additional signal field $ {\bf n}$, see Appendix~\ref{sect_path_integral}. Fig.~\ref{fig_instanton_in_marketron} shows the instanton solution obtained by numerical integration of the instanton equations \eqref{instanton_eq_gen} for the marketron potential \eqref{Langevin_potential}.

\begin{figure}[ht]
\begin{center}
\includegraphics[width=175mm,height=60mm]{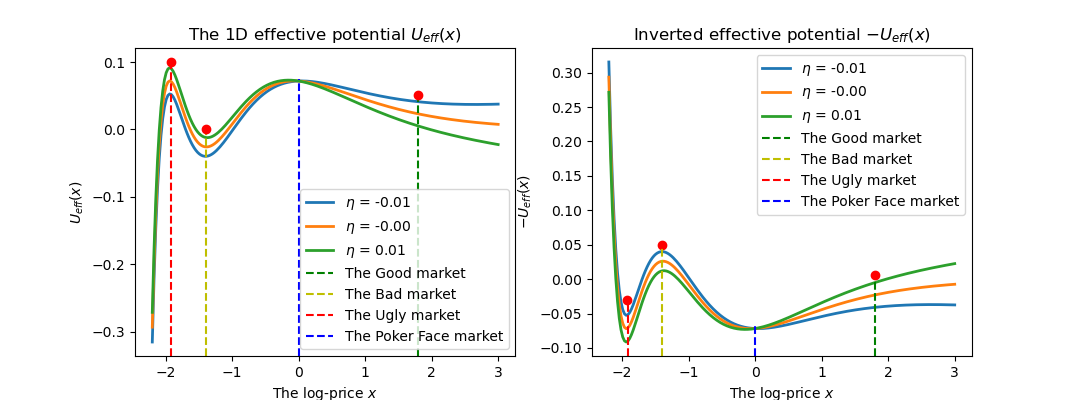}
\caption{On the left: $U_\eff(x)$ from Fig.~\ref{fig_Langevin_pot_1D_eff}. On the right: $-U_\eff(x)$.
Marketron parameters' values are: $c(t) = 0.13, g = 0.25, \mu = 0.1, \bar{y} = 1.$. While transitions between the Bad market and other market regimes are forbidden in classical mechanics for the original potential $U_\eff(x)$ (as they require extra energy), they become classically allowed for the inverted potential $-U_\eff(x)$.}
\label{fig_Inverted_potential}
\end{center}
\end{figure}

In our model, in addition to instantons obtained for the full 2D formulation, we can also analyze instantons arising in the simplified
one-dimensional formulation of the model obtained in the D-limit. The behavior of the instanton solution for the effective 1D potential \eqref{pot_eff} is illustrated in Fig.~\ref{fig_Inverted_potential}. The left plot depicts a classical particle, placed at the bottom of the local minimum of the original potential $U_\eff(x)$, which cannot go over the barrier to reach the state with the same energy. This is because getting there would require extra energy to go over the top of the barrier. Therefore, such a transition would be forbidden in classical mechanics and is only possible in statistical or quantum mechanics because they {\it do} allow for such energy fluctuations. However, if we flip the sign of the potential (the right plot), such motion becomes {\it classically allowed}, as it now corresponds to the particle sliding downward from the top of the hill. The top of the hill is located at the same position, which was the point of the minimum of the original potential on the left.  Appendix~\ref{sect_metastability_in_1D} provides an analytical approach to computing instanton transitions in the D-limit of our model where the dynamics become one-dimensional.

\begin{figure}[ht]
\begin{center}
\includegraphics[width=185mm,height=75mm]{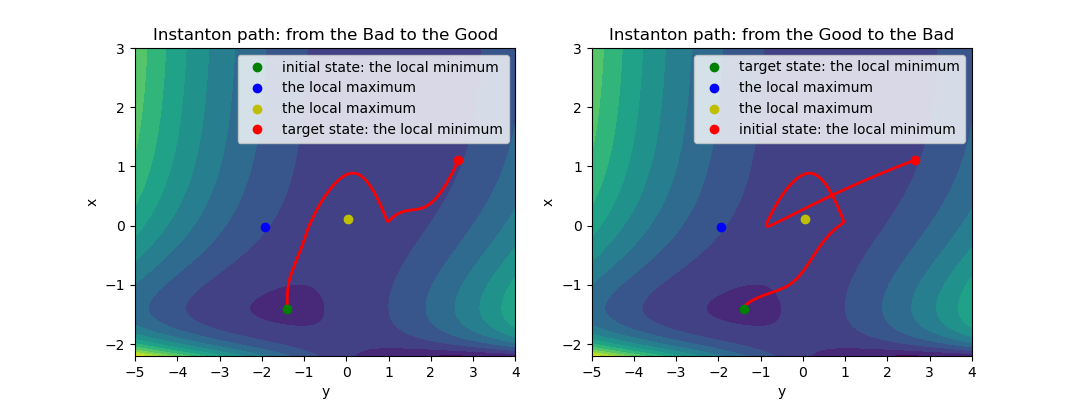}
\caption{Instanton solutions describing transitions from the Bad to the Good market (on the left), and from the Good to the Bad market (on the right).
}
\label{fig_instanton_in_marketron}
\end{center}
\end{figure}

\subsection{Metastability: market regimes, volatility clustering, crashes and defaultable equity}

The Langevin potential \eqref{Langevin_potential} assigns a non-zero probability to events where the stock prices experience large negative drops to very low values, or even to a strictly zero price level. To this end, the marketron, as described by the $x$-variable, should overcome the potential barrier in Fig.~\ref{fig_Inverted_potential} via an instanton. Depending on model parameters and the resulting height of the barrier, the probability of such an instanton transition can be made arbitrarily small. Getting to the edge of this regime, i.e., the peak of the potential barrier, means that the market at this point becomes the Ugly market.\footnote{As mentioned above, our model does not really predict or care whether the final state, reached after getting to this tipping point, is absorbing or not. We only care about the probability of getting to this point or, closely related to it, the probability of finding the particle on the other side of the barrier.} On top of this, the model suggests two other market regimes (the Bad and Good markets) that are similarly separated by potential barriers.

As a result, our model predicts three market regimes. Due to the barriers separating different regimes, the model dynamics in the vicinity of every barrier\footnote{Again, with a possible exception for the Ugly market's state, which can be made absorbing or non-absorbing at will, because, as mentioned in the previous comment, this decision would be outside of the model proper.} become {\it metastable}. This means that the potential barriers are sustainable only for a period of time whose expected length depends on the height of the barrier. Once in a while, the system quickly undergoes a transition into another metastable state, and so on. Note that the realized stock volatility will be impacted by the second derivative of the potential at the corresponding local minimum. This suggests that the well-known market phenomena such as volatility clustering and price-volatility correlation might have a natural explanation in our framework as transitions between the states of the metastable potential caused by a proper choice of the parameters of the marketron potential. It is also worth emphasizing that the effective potential itself is {\it random} as its shape is impacted by the signal $z_t$.

In addition to the Good and Bad markets, the Ugly market describes events of extreme jumps in market prices to the left (i.e., to decreasing $x$), which may go to arbitrarily low values including zero. Such events describe severe market crises, or even a complete market crash, if we apply this model to the market as a whole. On the other hand, when applied to individual stocks, it becomes a model of {\it defaultable equity}, with a non-zero instanton-induced probability for the stock price to move all the way to zero (or to negative infinity in the $x$-space).

As instantons proceed very quickly in time, they can be used to explain corporate defaults or market crashes without introducing additional exogenous state variables. Thus, instantons can describe events that really occur in the market, namely market crashes, corporate defaults, and bankruptcies, {\it without introducing new independent state variables}, e.g., a jump component in the price dynamics, hazard rates for additional Poisson processes, etc., that might drive such events. Furthermore, publicly tradable firms have 'market-implied' probabilities of default that can be expressed via market prices (market spreads) of credit default swaps referencing these firms. This dependence can be used to estimate parameters of our model using market data, see \cref{sect_model_estimation}.

The ability to capture default risk in equity prices and market crashes without introducing exogenous state variables is a valuable feature of our framework. This distinguishes our model from traditional approaches like the GBM model, where corporate default (when stock price $S$ reaches zero) is mathematically impossible, contradicting real-world observations. In our model, equity default risk emerges naturally from the relationship between market performance and money flows - specifically, that money tends to flow more readily into well-performing markets.

Conversely, when the cash supply to a stock can be arbitrarily high (large values of $y$), default becomes impossible, which aligns with the model's expected behavior. Both market crashes and defaults can be understood as instanton transitions through a potential barrier. When $g > 0$, this barrier is created by money flow into either an individual stock or the market as a whole, with its price impact determined by \eqref{quadr_impact}. Ultimately, this represents a flow-generated feedback mechanism where the dependency of money flows on market performance creates the barrier itself.

At the end, it is worth mentioning that in \cite{BC, QED} the Langevin approach was already used to explain market price fluctuations. For instance, in \cite{BC} the authors consider a nonlinear Langevin dynamics in the space of the stock price's speed, derived from the balance of the stock's demand and supply. With that approach, market crashes were interpreted as events of Kramer's escape (see, e.g., \cite{MELNIKOV19911} among others) caused by an imbalance of supply and demand. A similar Langevin dynamics, but in the log-return space and with a different nonlinear Langevin potential that produces instantons, was presented in \cite{QED}. The latter model constructed a nonlinear Langevin potential based on the analysis of money flows and their price impact (as in the present paper), but using different and less realistic parameterizations of both. While \cite{QED} demonstrated the possibility of finding explicit instanton solutions and calibrating the model to both stock prices and credit default swap spreads, our work develops these ideas further by introducing memory effects and multiple market regimes.

\section{Alternative views on marketron} \label{neuron-sect}

\subsection{Marketron as a spiking neuron of the market}

To summarize our development thus far, we have proposed two Langevin equations \cref{controlled_Langevin,Lang_y_2} to govern the marketron dynamics. For clarity, we rewrite these equations supplemented by a single OU process for the hidden signal $\theta_t$ (assuming for simplicity that in \eqref{OU} $K = 1$)
\begin{align} \label{Marketron_3D}
dx_t &= \left[ v f(\theta_t)  + \eta -  c(t) y_t V'_M(x) \right] dt + \sigma d W_t, \\
dy_t &= \left[ v h(\theta_t) + \mu (\bar{y} - y_t) -  c(t) V_{M}(x) \right] dt + \sigma_y d \tilde{W}_t, \nonumber  \\
d \theta_t &= k( \hat{\theta} - \theta_t) dt + \sigma_{z} d Z_{t}, \qquad x(0) = x_0, \quad y(0) = y_0, \quad \theta(0) = \theta_0, \nonumber
\end{align}
\noindent where all Brownian motions are assumed to be independent. This system of non-linear SDEs, coupled via nonlinear drift terms, provides a general formulation of the marketron model. While Markovian in the 3D space of all state variables $x_t, y_t, \theta_t$, the model becomes non-Markovian when considering only the $x$-variable, due to its incorporation of memory effects from past money flows and active forces (signals) $z_t$.

Most interestingly, it strongly resembles models of spiking neurons developed in neuroscience, such as the FitzHugh-Nagumo (FN) model \cite{FN, Tuckwell_1998}. The FN model describes a spiking neuron as a nonlinear oscillator whose spiking activity is controlled by a memory variable and an external stimulus.
Its two variables represent the voltage variable and the recovery variable associated with the concentration of potassium in the axon, which are analogous to our variables $ x $ and $ y $, respectively.

The FN model is known to produce a very rich range of interesting dynamic behavior in different parameter regimes, including in particular metastability and limit cycles \cite{Tuckwell_1998, Kurrer_1991, Stiefel2016}. We note that if we expand the exponents up to the quartic nonlinearity and approximate the interaction between the $x$- and $y$-oscillator by a linear term, the resulting model would be very similar to the stochastic FN model.

Our model thus produces dynamics of the marketron that are very similar to the dynamics of a spiking neuron in the FN model. We can think of our marketron as a {\it spiking neuron} (nonlinear oscillator) of the market, where a spike event would be associated with a large negative jump describing a market crash or corporate default, depending on the interpretation of the model. In our model, such jumps occur by the instanton mechanism. Moreover, their intensity is controlled by the memory variable $y_t$ that carries information about past flows, much like in the FitzHugh-Nagumo model. Based on its mathematical similarity to the latter, we can expect that our model can similarly produce a wide range of dynamic scenarios including both stable and metastable dynamics facilitated by both money flow $u_t$ and signals $z_t$.

\subsection{Market dynamics as controlled active matter} \label{sect_controlled_dynamics}

Returning to the general time-dependent controlled 2D Langevin dynamics in \cref{controlled_Langevin,Langevin_y}, recall that process $z_t$ describes "predictive signals" for market returns, represented as a sum of uncorrelated OU processes \eqref{OU}. Although this framework is common in the financial literature, where predictive signals typically derive from firms' fundamentals or past market prices, in this paper we propose an alternative interpretation drawing on methods from the physics of {\it active matter} \cite{Fodor_2022,Byrne_2022}.

Active matter consists of large numbers of active "agents" that consume energy to move or exert mechanical forces. Such systems are inherently out of thermal equilibrium. Active particles can represent various systems ranging from synthetic to living entities, such as swarming bacteria, microorganism collections, ant colonies, or bird flocks. These particles convert environmental energy into {\it self-propulsion}, sometimes producing collective effects without equilibrium analogs. Many active matter systems, particularly microorganisms and microswimmers, experience thermal fluctuations which give rise to diffusive components in their self-propelled motion. In addition, the self-propulsion mechanism is typically noisy itself, due to either environmental factors or intrinsic stochasticity of a self-propulsion mechanism.

A popular approach to modeling active matter is to use an overdamped Langevin equation similar to \eqref{controlled_Langevin}, where an OU process $z_t$ is used to capture a self-propulsion force acting on an active particle, in addition to a force due to the gradient of an external potential acting on the particle, and a Gaussian white noise which is added to handle thermal fluctuations.

Importantly, signals $z_t$ are typically modeled as {\it unobservable} OU processes, representing either environmental active components or particle self-propulsion mechanisms \cite{Dabelow_2019}. Furthermore, in problems that involve {\it control} of active particles, some parameters of a Langevin potential are assumed to be externally controlled to achieve certain objectives such as moving the system from one state to another at minimal cost or in minimal time.

Now let us go back to our Eqs.(\ref{Marketron_3D}). While we constructed these Langevin dynamics as a {\it financial} model, they are mathematically identical to the controlled dynamics of active OU particles subject to an external potential and a thermal bath. To have a complete analogy, we only need to replace the observed signals $z_t$ as commonly used in financial models with the {\it unobserved} OU processes that capture a self-propulsion property of stock prices. These processes can represent firms' entire production and corporate activities, jointly producing an active self-propulsion component in market price dynamics. This is simply because all such firms' actions can be viewed in our framework as activities trying to drive the market value of the firm's equity up. While an impact of such corporate efforts on market stock price is widely acknowledged, associating it with specific observable processes proves challenging, making the unobserved self-propulsion assumption natural.

Furthermore, similar to the setting of physics models where the 'active component' process is assumed to be independent of a control protocol \cite{Fodor_2023}, the same assumption applies in our financial model, as both production activities and corporate actions are reasonably independent of money flows to the stock market.\footnote{The stock market is a secondary market for equity issued by firms, therefore if we encode firms' activities into unobservable OU signals $z_t$, it appears reasonable to view them as independent of market flows, at least in a first approximation.}

\section{Model estimation} \label{sect_model_estimation}

This section presents calibration of the model in \eqref{Marketron_3D} to market data. Since the only observable variable in \eqref{Marketron_3D} is $x_t$, one approach to calibrate the model is to fit it to the time series of the S\&P500 index and use nonlinear filtering to capture the dynamics of unobservable variables $ y_t $ and $ \theta_t $. For a review of nonlinear filtering, see e.g. \cite{Daum2005,Simon2006,Setoodeh2022}.


Since this paper is focused on situations where the marketron potential has a form as in Fig.~\ref{fig_Inverted_potential} (where transitions between various regimes occur via the instanton mechanism), additional constraints should be imposed when performing filtering because not every calibration gives rise to this form of potential. In other words, for a given time series of index returns, there could be periods when parameters of the model found by calibration are such that the potential function doesn't exhibit three extrema. By adding an additional constraint as in \cref{appConstr}, we explicitly require the calibrated parameters to preserve the necessary shape. Obviously, this could potentially make convergence of the calibration worse, if not prevent it entirely. Nevertheless, our goal is to investigate whether convergence in constrained settings is achievable and how different the results are from those obtained with unconstrained filtering.


As shown in \cref{appConstr}, the additional (nonlinear) constraints that preserve the necessary shape of the marketron potential read
\begin{align} \label{finConstrRed}
0 &> J(t) [g I(t) - J(t)], \quad \Delta > 0, \\
0 &> 8 c(t) J(t) \left[ g I(t) + J(t) \right] + c(t)^2 \left[ g I(t)  + 2 J(t) \right]^3 - 8 g^2 \eta J(t), \nonumber \\
I(t) &= \left(\bar{y} + \frac{v h(\theta_*)}{\mu} \right) \left(1 - e^{- \mu t} \right) + y(0) e^{-\mu t}, \qquad
J(t) = c(t) \frac{1 - e^{- \mu t}}{\mu}, \nonumber
\end{align}
\noindent where $\Delta$ is defined in \eqref{finConstr}, and $\theta_*$ could be either $\theta_0$ (slow relaxation) or $\hat{\theta}$ (fast relaxation). These conditions are necessary for the potential to have four real roots. In addition, we append them with the relaxed conditions: $\Delta < 0$ for having two real roots, or $\Delta = 0, P < 0, D < 0, \Delta_0 \ne 0$ for having three real roots (again, see \eqref{finConstr} for the definition of $P, D, \Delta_0$). Thus, those are the constraints we impose on parameters of the marketron model when doing nonlinear filtering.

Note that $\mu > 0$ implies $J(t) > 0$. Therefore, the first and second constraints in \eqref{finConstrRed} become simplified.

\subsection{Calibration} \label{calib}

Since variables $\theta_t, y_t$ in \eqref{Marketron_3D} are unobservable, for calibration of this system we use filtering. The system in \eqref{Marketron_3D} is highly nonlinear due to nonlinearities in all drift terms, and, as we verified, simple filtering methods such as a maximum likelihood method combined with an extended Kalman filter, \cite{Date2010,DurbinKoopman2001}, don't produce satisfactory results. Therefore, we employ a particle filter method, see, e.g., \cite{Li2015} and references therein.

Let us consider a discrete time model and denote the estimate of $y_t$ at time $t_n$ as $y_n$, and $\theta_n$ as an estimate of $\theta_t$, respectively. Let us also denote the estimates of $y_n, \theta_n$ based on information up to time $t_{n-i}$ as $\hatE{y}{n}{n-i}, \hatE{\theta}{n}{n-i}$ for $i > 0$. We assume that the initial estimates $\hatEo{y}, \hatEo{\theta}$ are known. Next, assuming that nonlinearities in \eqref{Marketron_3D} are smooth, we expand the nonlinear drifts into Taylor series around $\hatE{y}{n}{n-1}, \hatE{\theta}{n}{n-1}$ as
\begin{align} \label{discr}
\Theta_d(t_n) &= k \left( \hat{\theta} - \hatE{\theta}{n}{n-1} \right), \\
Y_d(t_n) &= v h(\hatE{\theta}{n}{n-1}) + v h'(\hatE{\theta}{n}{n-1}) d \theta_n + \mu(\bar{y} - \hatE{y}{n}{n-1}) - [c(t_n) + c'(t_n) \Delta t] V_M(x_n), \nonumber \\
X_d(t_n) &= v f(\hatE{\theta}{n}{n-1}) + v f'(\hatE{\theta}{n}{n-1}) d \theta_n + \eta -[c(t_n) + c'(t_n) \Delta t] \hatE{y}{n}{n} V'_M(x_n), \nonumber
\end{align}
\noindent with
\begin{align}
d \theta_n &= k( \hat{\theta} - \hatE{\theta}{n}{n-1}) dt + \sigma_{z} \sqrt{\Delta t} w^{(\theta)}_{n}, \\
\hatE{y}{n}{n} &= \hatE{y}{n}{n-1} + \left[ v h(\hatE{\theta}{n}{n-1}) + v h'(\hatE{\theta}{n}{n-1}) d \theta_n + \mu (\bar{y} - \hatE{y}{n}{n-1}) -  c(t) V_{M}(x) \right] dt + \sigma_{y} \sqrt{\Delta t} w^{(y)}_{n}, \nonumber
\end{align}
\noindent where $\Delta t$ is the time step and $w^{(y)}_{n}, w^{(\theta)}_{n}$ are uncorrelated Brownian motions with zero mean and unit variance. In \eqref{discr} we use a substitution (Seidel's) scheme where new values of variables found in the first equation are immediately substituted into the second one, and so on.

Using the Euler-Maruyama scheme to discretize \eqref{Marketron_3D}, this system can be rewritten as
\begin{align} \label{Marketron_3D_d}
x_{n+1} &= x_n + X_d(t_n) \Delta t + \sigma \sqrt{\Delta t} w^{(x)}_{n}, \\
y_{n+1} &= \hatE{y}{n}{n-1} + Y_d(t_n) \Delta t + \sigma_y \sqrt{\Delta t} w^{(y)}_{n}, \nonumber \\
\theta_{n+1} &= \hatE{\theta}{n}{n-1} + \Theta_d(t_n) \Delta t + \sigma_z \sqrt{\Delta t} w^{(\theta)}_{n}, \qquad n=0,1,\ldots,N. \nonumber
\end{align}
In this system, unknown variables (the calibration parameters) are: $\sigma, \sigma_y, \sigma_z, v, \eta, k, \mu, g, \hat{\theta}, c(t), \bar{y}$, and the initial values $y(0), \theta(0)$ are set to zero.

It is worth noting that there are two sources of nonlinearity in \eqref{Marketron_3D}: i) the signals via the functions $f(\theta_t), h(\theta_t)$; and ii) a nonlinear drift of the observable variable $x_t$. A similar class of systems was considered in \cite{ChenLiLiu2022}, who claim that for this special class of nonlinear systems, closed analytic formulae for computing the conditional statistics could be obtained, which means that this system can be calibrated with no additional errors introduced by discretization. However, we didn't achieve much success along this path. On the other hand, a standard Matlab particle filter with the number of particles equal to 1500 produced reasonable results that will be presented below.

When running the particle filter, it is assumed that the ratio of the error between predicted and actual measurements follows a Gaussian distribution with zero mean and variance 0.05.

As mentioned, in \eqref{Marketron_3D_d} the only measurable stochastic variable is $x$. Examining the first line of \eqref{Marketron_3D_d} and considering the discretization in \eqref{discr}, we can express the total noise term in the equation for $x_{n+1}$ as
\begin{align}
\bar{w}^{(x)}_{n} &= \sigma {w}^{(x)}_{n} - c(t) V'M(x_n) \Delta t \sigma_y  {w}^{(y)}_{n} + \Delta t \sigma_z {w}^{(\theta)}_{n} \left[ k_{1,x} b_1 f'(\hatE{\theta}{n}{n-1})  - c(t) V'M(x_n) k_{1,y} b_2 h'(\hatE{\theta}{n}{n-1})\right].
\end{align}
It can be seen that several calibration parameters enter this expression as products, creating an ambiguity, where different parameter values can yield the same product. To resolve this, we note that $\bar{w}^{(x)}_{n}$ is a martingale. Therefore, when running the optimizer, we impose additional "martingality" constraints requiring that the expectation (mean) of $\bar{w}^{(x)}_{n}$ over all paths (particles) must be zero. This additional constraint helps eliminate the parameter ambiguity in the solution.

\subsection{Implementation and results}

We calibrate our model to S\&P500 monthly log-prices from January 2000 to October 2024 obtained by aggregating daily time series.
\begin{figure}[!h]
\begin{center}
\hspace*{-0.5cm}
\subfloat[]{\includegraphics[width=0.55\textwidth]{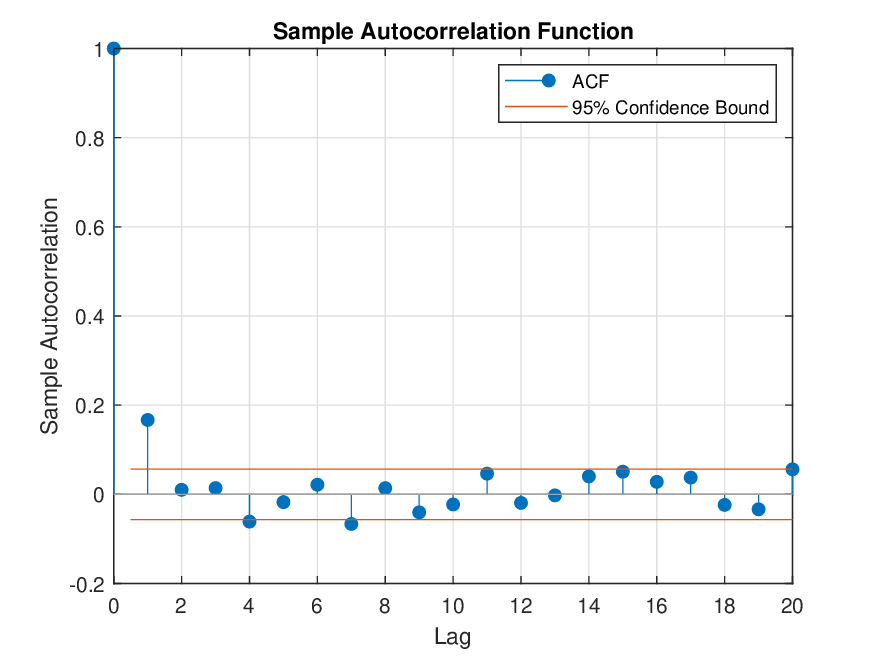}}
\hspace*{-0.7cm}
\subfloat[]{\includegraphics[width=0.55\textwidth]{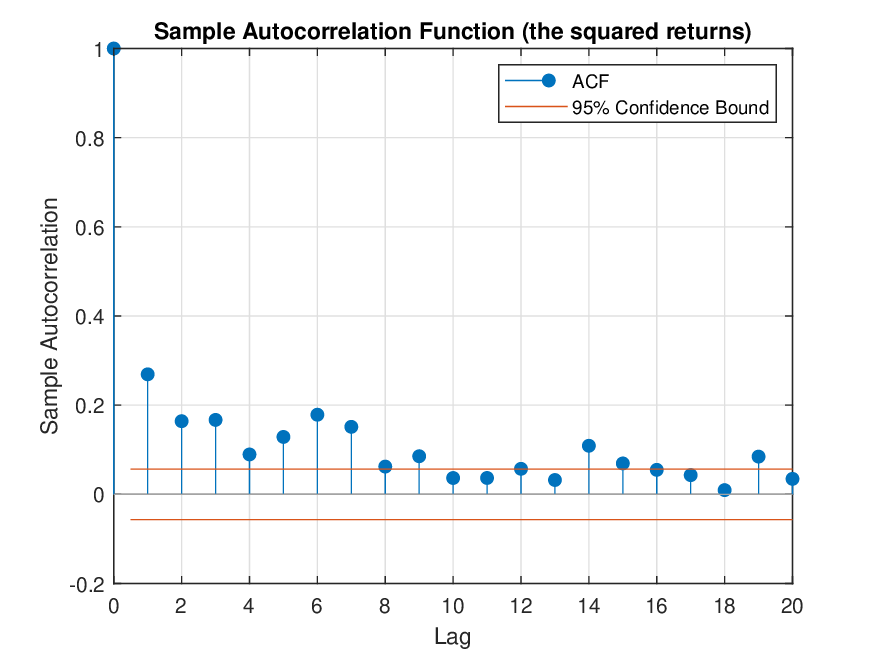}}
\end{center}
\vspace{-1em}
\caption{Autocorrelation of a) S\&P500 daily returns, and b) the squared returns.}
\label{autocorrelation}
\end{figure}
Fig.~\ref{autocorrelation} depicts the correlogram of the daily returns and squared returns computed using this time series. The serial correlation is small for all lags except lag one. The mean correlation is close to zero, and the correlation does not show any significant nonrandom variations. Therefore, returns can be approximately treated as i.i.d., which increases confidence in filter performance. However, the correlogram of the squared returns shows persistent serial correlation, which indicates that volatility clustering exists in the returns.

To recall, as the initial state we use $x_0 = x(0), \theta_0 = y(0) = 0, y_0 = y(0)$, where $x(0) = \log(S_0/S_*)$. We also set $\varepsilon = 0.02, S_* = 1000$.

To proceed, we need to explicitly define functions $f, h$. A detailed analysis of the first equation in \eqref{Marketron_3D} reveals that if $v f(\theta) + \eta < 0$, the log-price $x_t$ approaches the negative infinity as $ t \rightarrow \infty $, while in the opposite case it asymptotically tends to the positive infinity.

Both scenarios should have a very low probability within the time horizons considered in this paper. While such behavior would be expected for the 'true' set of model parameters in a realistic model, these parameters are initially unknown and must be determined through calibration. This creates a potential challenge for our calibration method, which relies on simulated trajectories and non-linear filtering. With arbitrary parameter values, an excessive number of trajectories tend to diverge toward negative or positive infinity. To address this issue, we introduce a more general specification of the signals using inhomogeneous functions $f, h$ that can change sign over time. These functions are defined as follows

\begin{align} \label{fhparam}
f(\theta) = a_1(t)/\left(1 + e^{-b_1 \theta} \right), \qquad h(\theta) = a_2(t)/\left(1 + e^{-b_2 \theta} \right), \\
a_1(t) = k_{1,x} \cos(k_{2,x} + k_{3,x} t), \qquad a_2(t) = k_{1,y} \sin(k_{2,y} + k_{3,y} t), \nonumber
\end{align}
\noindent where $b_1, b_2, k_{i,x}, k_{i,y}, , i \in [1,3]$ are constants determined by calibration. With this parameterization, our problem becomes overdetermined as the parameter $v$ enters the problem only in combinations $v a_1(t), v a_2(t)$. Therefore, to resolve this, we set $v=1$. Also, in what follows we assume $ c(t) $ does not change with time, i.e. $c(t) = c$. Thus, the total number of parameters to be found by calibration is 18.

To determine all model parameters, we formulate an optimization problem. Starting with initial parameter values, we run the particle filter to determine all predicted states of the model for the given horizon $H$. Then, the first four moments are computed using monthly time series constructed from predicted measurements over the interval $t \in [0,H]$. These moments are matched to those computed from market data, forming a least-squares objective function using the corresponding residuals. Finally, we combine objective functions across multiple horizons $H = [2,5,10,15,20,24]$ years into a single objective function. The Matlab package \verb'CEopt', \cite{CunhaJr2024CEopt} is used to minimize this objective function subject to the constraints in \eqref{finConstrRed}.

We find that \verb'CEopt' reaches an optimal solution in approximately 500 iterations, with the standard deviation error decreasing from 1.4e4 to 6.5. For such a run, we don't utilize a parallel version of the optimizer, and the elapsed time of our "sequential" calibration is about 1.5 hours. Given the stochastic nature of the cross-entropy algorithm, multiple calibration runs are feasible. To ensure reproducibility, we fix the random number generator seeds for both the particle filter and the CEopt algorithm.

Since the \verb'CEopt' solution might correspond to a local extremum despite its global search approach, we use it as an initial guess for a second global solver. We employ scipy differential evolution with the \verb'best2exp' mutation strategy and \verb'sobol' population initialization (see review in \cite{DE2011}). This solver addresses the same problem and constraints as \verb'CEopt'. It converges after approximately 200 iterations, achieving slightly better results (a lower objective function value) than \verb'CEopt', typically differing by about 10 percent. This supports our hypothesis that \verb'CEopt' converges to a local minimum while suggesting that this local minimum is relatively close to the global/local minimum found by \verb'CEopt'.

\subsubsection{The case $\theta_* = \theta_0 = 0$}

The parameters of the model found by this two-step calibration, together with the box constraints issued on them, are given in Table~\ref{calibParam}.
\begin{table}[tbhp]
\begin{center}
\scalebox{0.87}{
\begin{tabular}{|l|r|r|r|r|r|r|r|r|r|}
\toprule
\rowcolor[rgb]{ .792,  .929,  .984}
{\bf parameter} & $\bm \sigma$ & $\bm \sigma_y$ & $\bm \sigma_z$ & $\bm \eta$ & $\bm k$ & $\bm \mu$ & $\bm g$ & ${\bm \hat{\theta}}$ & ${\bm \bar{y}}$ \\ \hline
lower bound & 0 & 0 & 0 & $-\sigma_{\max}^2/2$ & 0 & 0 & 0 & 0 & 0 \\ \hline
value       & 0.7912 & 0.3800 & 0.8334 & -1.5685 & 1.2869 & 1.6671 & 0.6831 & 6.7865 & 0.4731  \\ \hline
upper bound & 3.0 & 3.0 & 3.0 & $1 - \sigma_{\min}^2/2$ & 5 & 3 & 1 & 10 & 1 \\ \hline
\rowcolor[rgb]{ .792,  .929,  .984}
{\bf parameter} & $\bm c$ & $\bm b_1$ & $\bm b_2$ & $\bm k_{1,x}$ & $\bm k_{2,x}$ & $\bm k_{3,x}$ & $\bm k_{1,y}$ & $\bm k_{2,y}$ & $\bm k_{3,y}$ \\ \hline
lower bound & 0 & -10 & -10 & -5 & 0 & -5 & -5 & 0 & -5 \\ \hline
value       & 3.9305 & 1.6819 & -1.2102 & -3.2002 & 2.7417 & -1.8832 & -0.7855 & 3.8901 & 1.5588 \\ \hline
upper bound & 5 & 10 & 10 & 5 & 5 & 5 & 5 & 5 & 5 \\
\bottomrule
\end{tabular}
}
\caption{Parameters of the model in \eqref{Marketron_3D} with the constraints in \eqref{finConstrRed} found by calibration to S\&P500 weekly returns from 2000 to Sept. 2024, together with the box constraints used in the calibration procedure.}
\label{calibParam}
\end{center}
\end{table}
Although our particle filter uses a small number of particles, it produces some statistics which can be analyzed to obtain more information about the dynamic behavior of the model. For instance, in Table~\ref{Stats} first four moments of the computed distribution of the log-returns are compared with those obtained by using our set of the market data.
\begin{table}[!htb]
\centering
\scalebox{0.87}{
\begin{tabular}{|r|r|r|r|r|r|r|r|r|}
\toprule
\rowcolor[rgb]{ .58,  .863,  .973} \multicolumn{1}{|c|}{\textbf{hor, yrs}} & \multicolumn{1}{c|}{\textbf{mnMarket}} & \multicolumn{1}{c|}{\textbf{mnModel}} & \multicolumn{1}{c|}{\textbf{stMarket}} & \multicolumn{1}{c|}{\textbf{stModel}} & \multicolumn{1}{c|}{\textbf{skMarket}} & \multicolumn{1}{c|}{\textbf{skModel}} & \multicolumn{1}{c|}{\textbf{kuMarket}} & \multicolumn{1}{c|}{\textbf{kuModel}} \\ \hline
 \rowcolor[rgb]{ .58, .863, .973} \textbf{2} &  \cellcolor[rgb]{ .71, .902, .635} -0.2002 &  \cellcolor[rgb]{ 1, 1, 1} -0.1854 &  \cellcolor[rgb]{ .71, .902, .635} 0.1962 &  \cellcolor[rgb]{ 1, 1, 1} 0.1969 &  \cellcolor[rgb]{ .71, .902, .635} -0.1640 &  \cellcolor[rgb]{ 1, 1, 1} -0.1602 &  \cellcolor[rgb]{ .71, .902, .635} 0.1797 &  \cellcolor[rgb]{ 1, 1, 1} 0.1779 \\ \hline
 \rowcolor[rgb]{ .58, .863, .973} \textbf{5} &  \cellcolor[rgb]{ .71, .902, .635} -0.0615 &  \cellcolor[rgb]{ 1, 1, 1} -0.0596 &  \cellcolor[rgb]{ .71, .902, .635} 0.1853 &  \cellcolor[rgb]{ 1, 1, 1} 0.2021 &  \cellcolor[rgb]{ .71, .902, .635} -0.1673 &  \cellcolor[rgb]{ 1, 1, 1} -0.1716 &  \cellcolor[rgb]{ .71, .902, .635} 0.1800 &  \cellcolor[rgb]{ 1, 1, 1} 0.1904 \\ \hline
 \rowcolor[rgb]{ .58, .863, .973} \textbf{10} &  \cellcolor[rgb]{ .71, .902, .635} -0.0441 &  \cellcolor[rgb]{ 1, 1, 1} -0.0402 &  \cellcolor[rgb]{ .71, .902, .635} 0.2007 &  \cellcolor[rgb]{ 1, 1, 1} 0.2232 &  \cellcolor[rgb]{ .71, .902, .635} -0.3131 &  \cellcolor[rgb]{ 1, 1, 1} -0.3098 &  \cellcolor[rgb]{ .71, .902, .635} 0.3988 &  \cellcolor[rgb]{ 1, 1, 1} 0.4068 \\ \hline
 \rowcolor[rgb]{ .58, .863, .973} \textbf{15} &  \cellcolor[rgb]{ .71, .902, .635} 0.0429 &  \cellcolor[rgb]{ 1, 1, 1} 0.0403 &  \cellcolor[rgb]{ .71, .902, .635} 0.1830 &  \cellcolor[rgb]{ 1, 1, 1} 0.2000 &  \cellcolor[rgb]{ .71, .902, .635} -0.3469 &  \cellcolor[rgb]{ 1, 1, 1} -0.3470 &  \cellcolor[rgb]{ .71, .902, .635} 0.4342 &  \cellcolor[rgb]{ 1, 1, 1} 0.4593 \\ \hline
 \rowcolor[rgb]{ .58, .863, .973} \textbf{20} &  \cellcolor[rgb]{ .71, .902, .635} 0.0704 &  \cellcolor[rgb]{ 1, 1, 1} 0.0632 &  \cellcolor[rgb]{ .71, .902, .635} 0.1693 &  \cellcolor[rgb]{ 1, 1, 1} 0.1811 &  \cellcolor[rgb]{ .71, .902, .635} -0.3518 &  \cellcolor[rgb]{ 1, 1, 1} -0.3657 &  \cellcolor[rgb]{ .71, .902, .635} 0.4620 &  \cellcolor[rgb]{ 1, 1, 1} 0.5232 \\ \hline
 \rowcolor[rgb]{ .58, .863, .973} \textbf{24} &  \cellcolor[rgb]{ .71, .902, .635} 0.0871 &  \cellcolor[rgb]{ 1, 1, 1} 0.0886 &  \cellcolor[rgb]{ .71, .902, .635} 0.1773 &  \cellcolor[rgb]{ 1, 1, 1} 0.1788 &  \cellcolor[rgb]{ .71, .902, .635} -0.3971 &  \cellcolor[rgb]{ 1, 1, 1} -0.3334 &  \cellcolor[rgb]{ .71, .902, .635} 0.4993 &  \cellcolor[rgb]{ 1, 1, 1} 0.4809
\\ \bottomrule
\end{tabular}%
}
\caption{Annualized statistics of log-returns produced by the market and the marketron model \eqref{Marketron_3D} with the model parameters found by calibration. Here {\bf mn, st, sk, ku} denote mean, volatility, skewness and kurtosis, respectively, and {\bf hor} is the horizon in years.}
\label{Stats}
\end{table}%
It can be seen that for all horizons, the log-returns exhibit a negative skew and positive kurtosis, while daily returns are normally distributed. This fact was already discussed in the literature, see e.g \cite{NeubergerPayne2019} and references therein, where the authors apply a proxy technique to U.S. stock index returns and show that skew is large and negative and does not significantly attenuate with horizon as one moves from monthly to multi-year horizons. We thus observe that the marketron model is capable of replicating the sign of the skewness and kurtosis.
\begin{table}[!htb]
\centering
\begin{tabular}{|r|r|r|r|r|}
\toprule
\rowcolor[rgb]{ .792,  .929,  .984} \multicolumn{1}{|c|}{\textbf{horizon, yrs}} & \multicolumn{1}{c|}{\textbf{mean}} & \multicolumn{1}{c|}{\textbf{volatility}} & \multicolumn{1}{c|}{\textbf{skewness}} & \multicolumn{1}{c|}{\textbf{kurtosis}} \\ \hline
2  &  2.7567 &  0.3699 & -0.0161 &  0.0699 \\ \hline
5  &  0.9400 &  0.4263 &  0.1615 &  0.1249 \\ \hline
10 &  1.3349 &  0.6174 &  0.0670 &  0.0993 \\ \hline
15 &  2.2060 &  0.7377 &  0.1640 &  0.1489 \\ \hline
20 &  4.6837 &  1.1648 &  0.1176 &  0.0994 \\ \hline
24 &  7.0418 &  1.5793 &  0.1091 &  0.0943 \\
\bottomrule
\end{tabular}%
\caption{Annualized statistics of the memory variable $y_t$ computed by using the model \eqref{Marketron_3D} with the model parameters found by calibration.}
\label{StatsY}%
\end{table}%
\begin{figure}[!htb]
\begin{center}
\includegraphics[width=\textwidth]{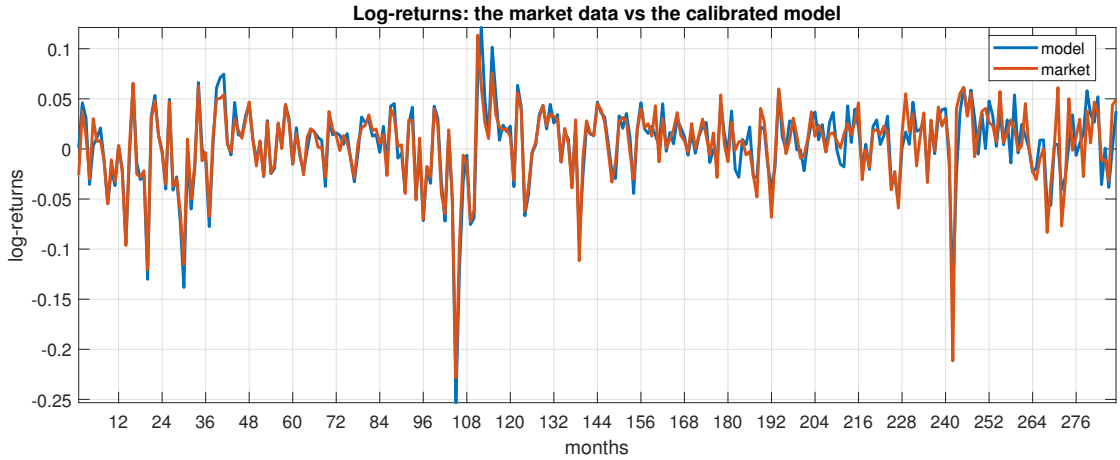}
\end{center}
\vspace{-1em}
\caption{Plot of log-returns computed by using the market data and by the marketron model with the model parameters found by calibration.}
\label{marketVsModelReturns}
\end{figure}
Though the impact variable $y_t$ is hidden in the sense that it is not directly observed in the market, it is instructive to generate similar statistics for $y_t$. These are presented in Table~\ref{StatsY}. It can be seen that the annualized skewness of $y_t$ is relatively small except at the horizon of 5 years, while small annualized kurtosis can be observed for almost all horizons. Thus, the distribution of $y_t$ is approximately Gaussian. However, the mean of the distribution is not zero and significantly changes with time, also reverting the sign.

\begin{figure}[!h]
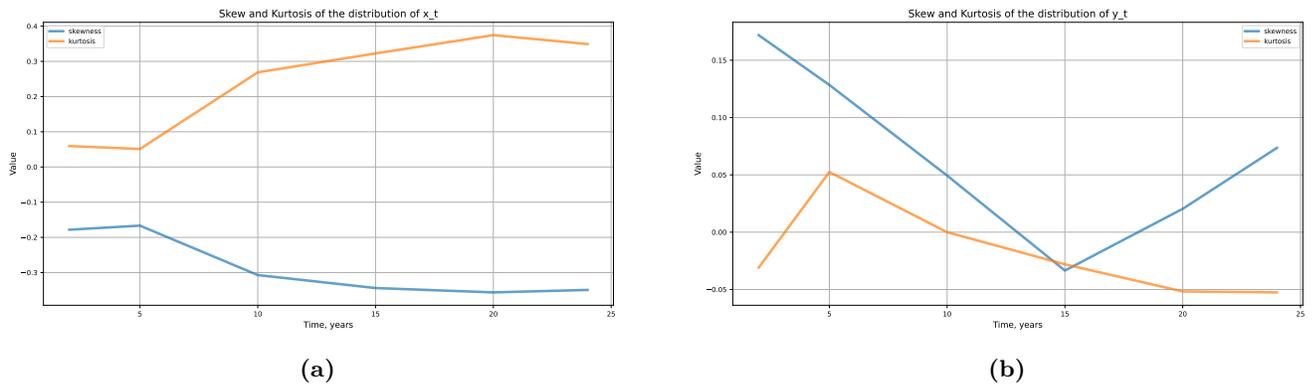

\begin{center}
\hspace*{-0.5cm}
\subfloat[]{\includegraphics[width=0.55\textwidth]{skewKurtosisX.pdf}}
\hspace*{-0.7cm}
\subfloat[]{\includegraphics[width=0.55\textwidth]{skewKurtosisY.pdf}}
\end{center}
\vspace{-1em}
\caption{Skewness and kurtosis as functions of time with the calibrated marketron model for a) log-returns $ x_t$, and b) the memory variable $y_t$.}
\label{skewKurt}
\end{figure}

Fig.~\ref{marketVsModelReturns} compares log-returns produced by the calibrated model with those observed in the market. Although the particle filter is run with a relatively small number of particles, the model parameters found by such calibration allow the model to approximately replicate the observed market data on S\&P500 monthly log-returns without significant outliers. The skewness and kurtosis of log-returns as functions of time are shown in Fig.~\ref{skewKurt}. The log-returns demonstrate a negative skew that decreases with time, while kurtosis shows an opposite behavior.

To further analyze the dynamic behavior of the model, we implemented a simple Euler-Maruyama Monte Carlo scheme to solve equation \eqref{Marketron_3D}, simulating $N = 10,000$ paths using the calibrated model parameters. Our analysis reveals that approximately 450 paths out of $N$ paths result in defaults which correspond to the escape to $x \to -\infty$ in the log-price $x$ space. This yields the annualized default intensity of approximately 18 bps, which aligns well with market-implied default intensities ranging from 10 to 50 bps based on credit market data.
%
\graphicspath{{./theta0/}}

\paragraph{Distributions of the model variables.} The next set of plots presents distributions of variables $x_t, y_t$ obtained in the simulation.
\begin{figure}[!htb]
\begin{center}
\scalebox{0.7}{
\fbox{\includegraphics[width=1.1\textwidth]{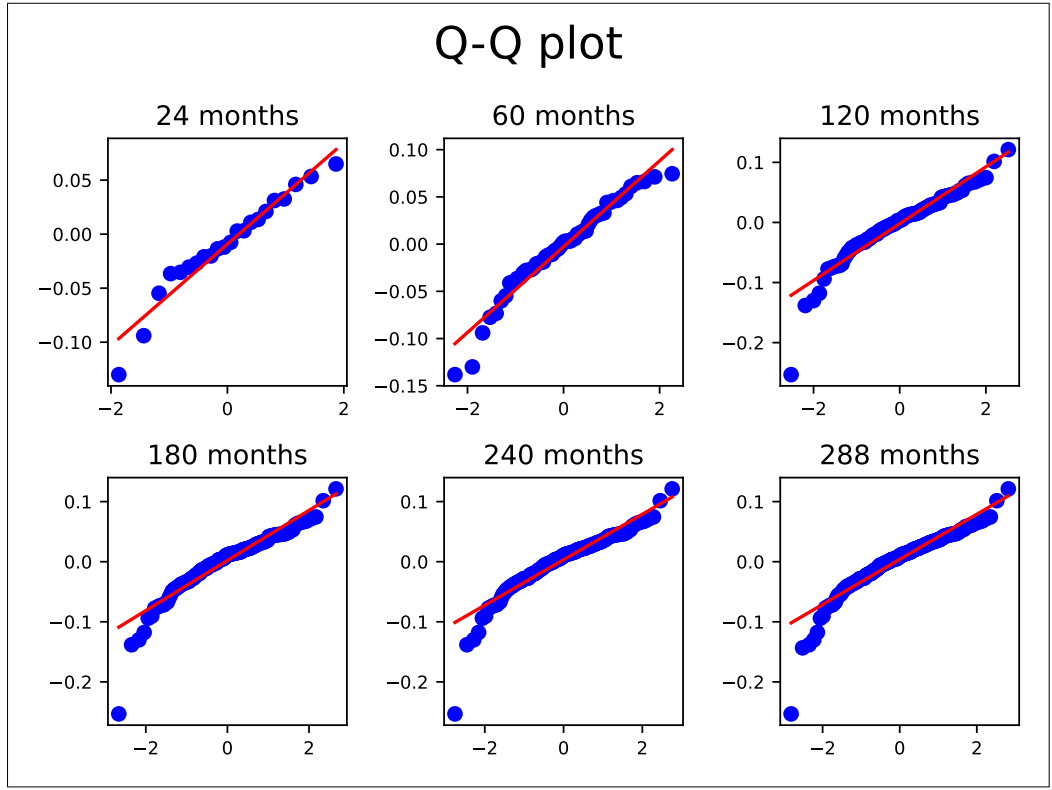}}
}
\end{center}
\vspace{-1em}
\caption{The $Q$-$Q$ plots of $x_t$ computed vs the normal distribution for different time horizons, obtained by simulation.}
\label{QQ}
\end{figure}
\begin{figure}[!htb]
\begin{center}
\fbox{\includegraphics[width=0.7\textwidth]{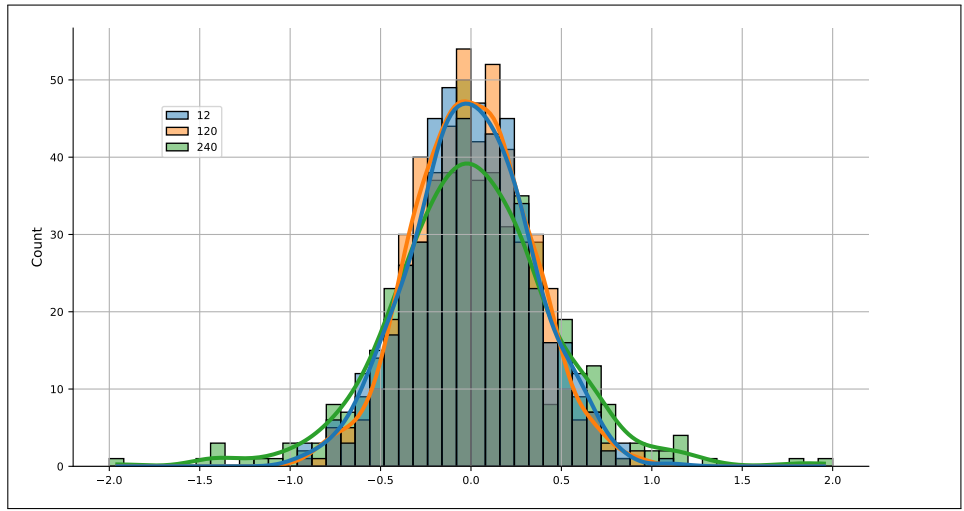}}
\end{center}
\caption{The distributions of the log-prices $x$ at $t=1,10,20$ years obtained by simulation.}
\label{distrib}
\end{figure}
\begin{figure}[!htb]
\begin{center}
\fbox{\includegraphics[width=0.7\textwidth]{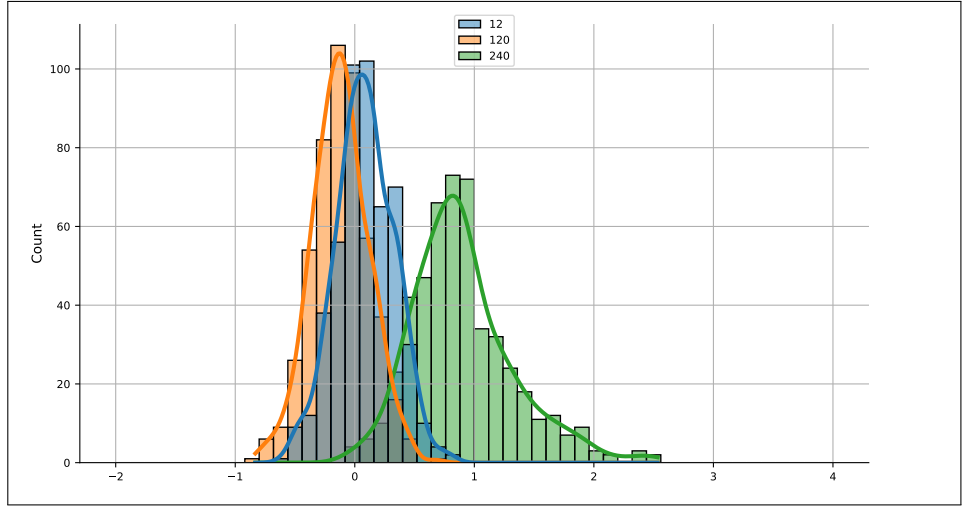}}
\end{center}
\caption{The distributions of the memory variable $y$ at $t=1, 10, 20$ years, obtained by simulation.}
\label{distribY}
\end{figure}

Fig.~\ref{QQ} presents Q-Q (quantile-quantile) plots comparing $x_t$ against the normal distribution across different time horizons. The analysis reveals distinct patterns of behavior at various time scales. At $t=1$ year, the log-prices follow an approximately normal distribution. As the horizon extends to 5 years, both tails begin to deviate from normality. Subsequently, while the right tail converges back toward the normal distribution, the left tail exhibits persistent deviation. This temporal evolution indicates that the distribution of log-prices becomes increasingly skewed over longer time horizons.

Fig~\ref{distrib} shows the distributions of the log-prices $x$ at $t=1, 10, 20$ years. The distributions are slightly {\it right-skewed}, but remain almost symmetric. The distributions of $y_t$ for $t=1, 10, 20$ years are presented in Fig~\ref{distribY}. Here, the {\it right skew} of all curves is more pronounced compared with the previous graph. It is interesting to note that while log-returns exhibit negative skewness, the log-prices are right-skewed.

Thus, nonlinearities in the drift for variables $x_t$ and $y_t$ give rise to skewed distributions of these variables. Therefore, the results obtained with the full marketron model qualitatively appear to behave similarly to those obtained with the piecewise harmonic approximation of the marketron potential, which are discussed in \cref{sect_approximate_potential}. However, in the latter case, a skewed BM is produced by weak singularities in the drift term, while here they are the consequence of a smooth drift nonlinearity.

\begin{figure}[!htp]
\begin{center}
\fbox{\includegraphics[width=0.8\textwidth]{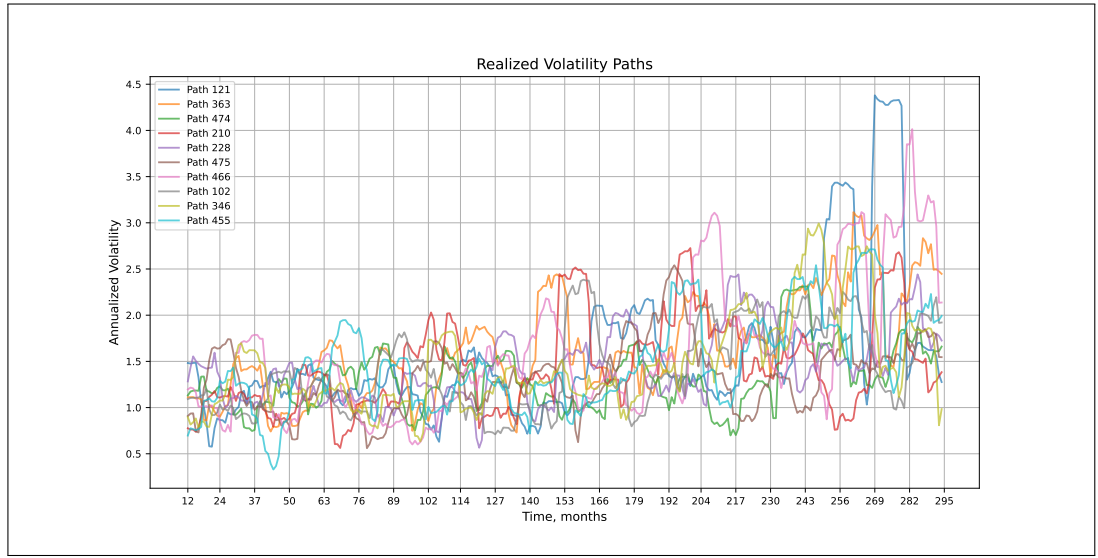}}
\end{center}
\vspace{-1em}
\caption{Annualized realized volatility of log-prices $x_t$ along several paths, obtained by simulation of \eqref{Marketron_3D} with calibrated parameters.}
\label{Paths}
\end{figure}

\begin{figure}[!htp]
\begin{center}
\fbox{\includegraphics[width=0.8\textwidth]{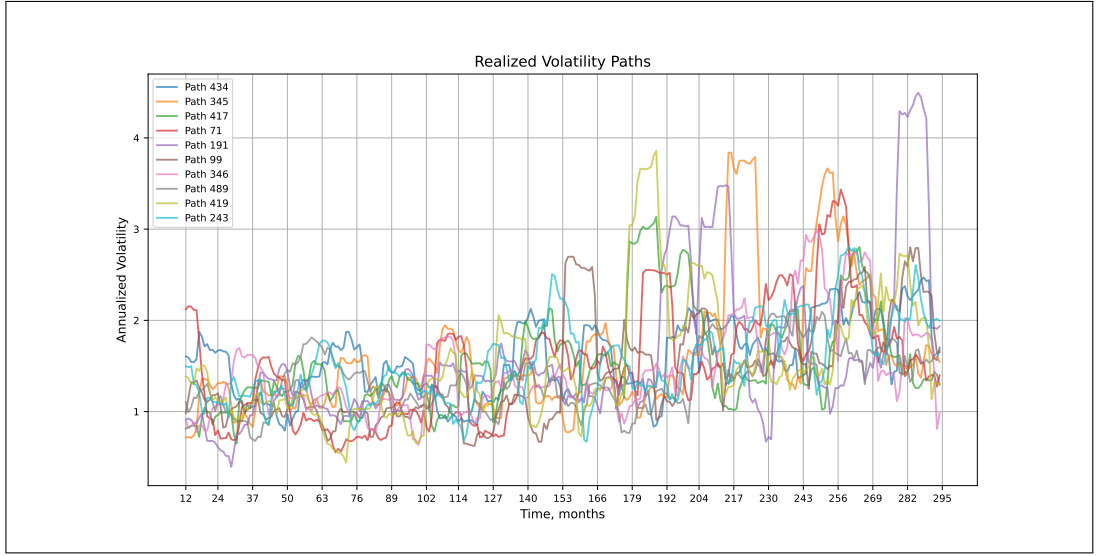}}
\end{center}
\vspace{-1em}
\caption{Annualized realized volatility of $y_t$ along several paths, obtained by simulation of \eqref{Marketron_3D} with calibrated parameters.}
\label{PathsY}
\end{figure}

\paragraph{The annualized realized volatility.} Fig.~\ref{Paths} shows the annualized realized volatility of the log-price as a function of time along several paths. Note that some time segments, e.g., around 2008, and especially from 2020 to 2024, might be associated with volatility clustering. A similar plot of annualized realized volatility of $y_t$ is presented in Fig.~\ref{PathsY}. Here, volatility clustering might also be observed in the right part of the graph corresponding to the recent data.

To quantify these observations, we computed the Hurst exponent $H$ of the volatility of log-returns obtained in our experiment. In agreement with the results in Fig.~\ref{QQ}, we found $H = 0.61$ at the left end of the series and $H = 0.44$ at the right end. These values indicate weak volatility clustering at the left end of the series and some roughness of the realized volatility  at the right end. It is important to note that since we used monthly time-series data, the series length is relatively short. For short time series, traditional algorithms for calculating the Hurst exponent typically have lower accuracy \cite{Nikolova2020}.

\begin{figure}[!htb]
\begin{center}
\scalebox{0.8}{
\subfloat[]{\includegraphics[width=0.55\textwidth]{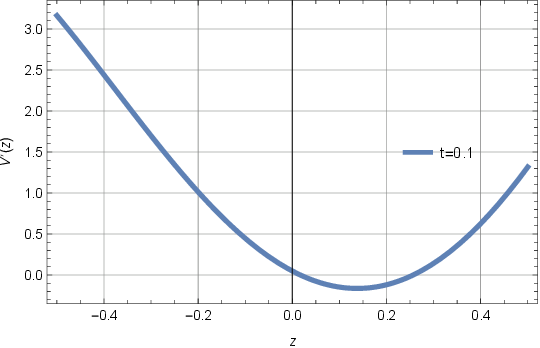}}
\subfloat[]{\includegraphics[width=0.55\textwidth]{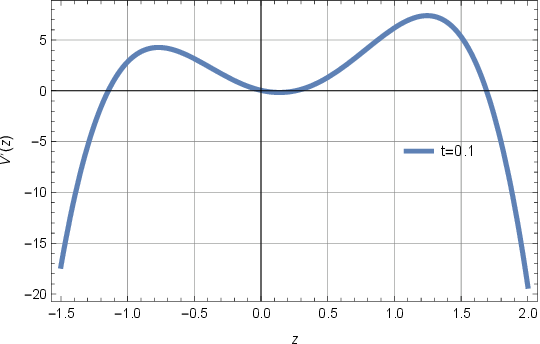}}
}
\end{center}
\vspace{-1em}
\caption{Plot of $z V'(z)$ as defined in \eqref{potVz} at $t = 0.1$: a) the zoomed-in picture to see two real roots close to the origin; b) the other two roots. }
\label{Vz-t_0.1}
\end{figure}

\begin{figure}[!htb]
\begin{center}
\scalebox{0.8}{
\subfloat[]{\includegraphics[width=0.55\textwidth]{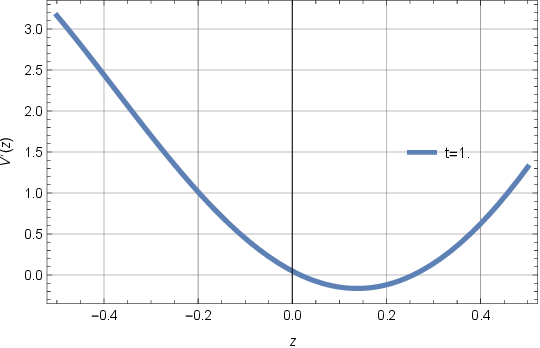}}
\subfloat[]{\includegraphics[width=0.55\textwidth]{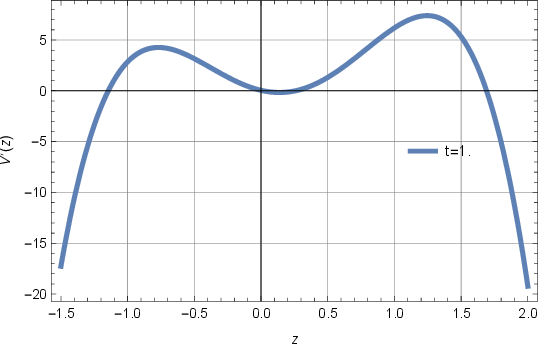}}
}
\end{center}
\vspace{-1em}
\caption{Plot of $z V'(z)$ as defined in \eqref{potVz} at $t = 1$: a) the zoomed-in picture to see two real roots close to the origin; b) the other two roots. }
\label{Vz-t_1}
\end{figure}

\paragraph{The marketron potential shape.}  To ensure these parameters produce the marketron potential of the expected shape, we plot the function $z V'(z)$ (as defined in \eqref{potVz}) by substituting the calibrated parameters of the model into the RHS of \eqref{potVz}.  Figs.~\ref{Vz-t_0.1} and ~\ref{Vz-t_1} show the behavior of this function for $t = 0.1$ and $t = 1$, respectively.
\graphicspath{{./Figs/}}
\graphicspath{{./theta0/}}
\begin{figure}[!htb]
\captionsetup[subfloat]{captionskip=-30pt}
\vspace{-0.4in}
\begin{center}
\scalebox{0.8} {
\hspace*{-0.9in}
\subfloat[]{\includegraphics[width=0.7\textwidth]{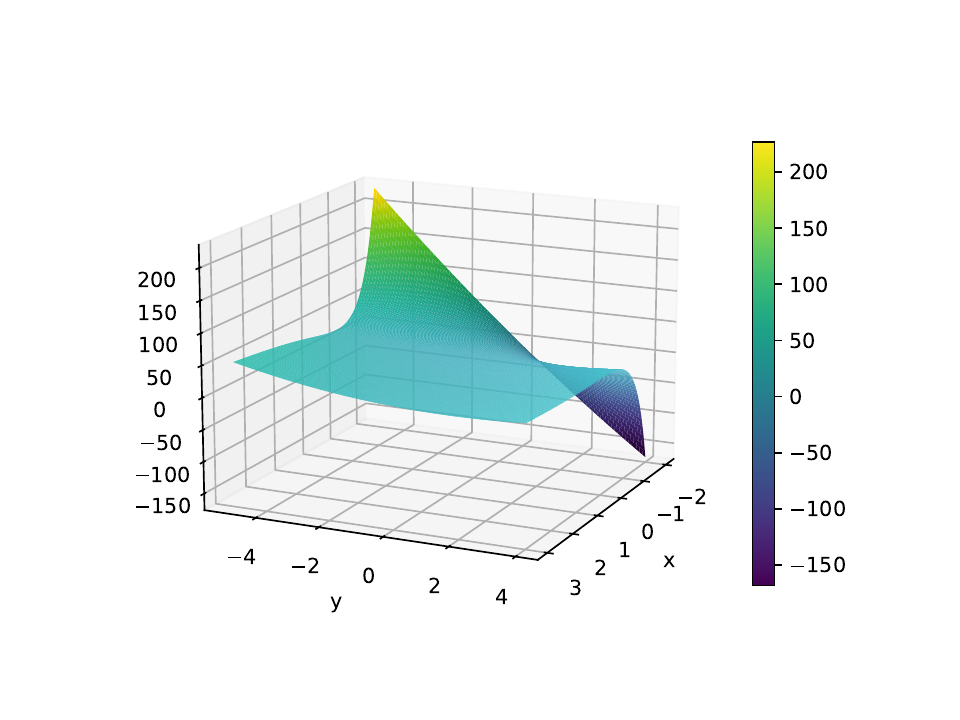}}
\hspace*{-0.4in}
\subfloat[]{\includegraphics[width=0.7\textwidth]{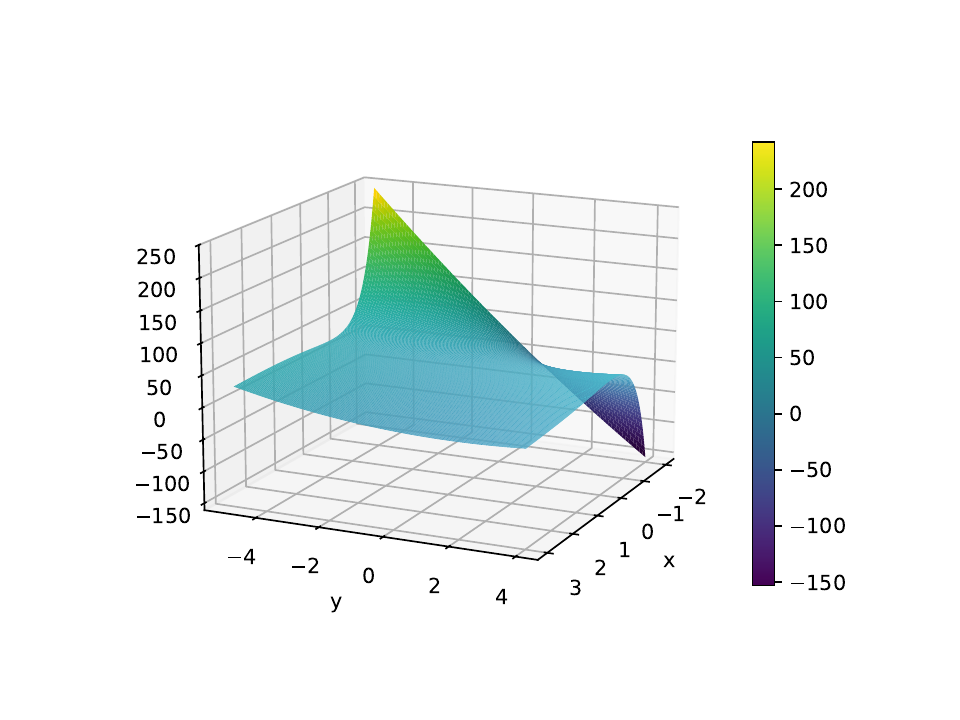}}
}
\end{center}
\vspace{-0.3in}
\caption{3D marketron potential $V(x,y)$ in \eqref{Lang_pot_general} computed with the model parameters found by calibration for a) $t=1$ year, b) $t=20$ years.}
\label{pot3Dcalib}
\end{figure}
\captionsetup[table]{skip=15pt}

It can be seen in Fig.~\ref{Vz-t_0.1} that the quartic polynomial $z V'(z)$ has four real roots. Two of them are positive and located close to the origin. The third root is also positive, while the fourth one is negative. Since $z = e^{-x}$ (as defined in \eqref{potVx}) is always nonnegative, this last root must be excluded. Thus, the derivative of the marketron potential in $z$ (or in $x$ since $z = z(x)$) has three real roots. Accordingly, the potential has three extrema, i.e., exactly what we tried to achieve.

Similar results at $t = 1$ are presented in Fig.~\ref{Vz-t_1}. Note that by the definition of $z$, the corresponding potential in $x$ will be inverted.

Fig.~\ref{pot3Dcalib} shows a 3D marketron potential $V(x,y)$ in \eqref{Lang_pot_general} computed with the model parameters found by calibration for times $t=1, 20$ years. In agreement with \cref{Vz-t_0.1,Vz-t_1} and the discussion in \cref{sect_The_Good_Bad_Ugly}, these results can be interpreted as the S\&P500 market being the Good market except in regions corresponding to large $x$ (or small $z$).

\paragraph{Predictive power of the model.} From a practical perspective, one may want to explore the predictive power of our model for short time horizons. While a three-month rolling window could ideally be used to compare the first four moments of log-returns between the model and market data, this approach is not feasible when operating with monthly log-returns that we work with in this paper.  As our calibrator uses monthly market log-prices, this would provide only three data points in the rolling window, which is obviously insufficient for a meaningful analysis. On the other hand, as discussed in more details in \cref{sect_Summary}, calibrating our algorithm using daily log-prices is computationally prohibitively expensive.
\begin{figure}[!htb]
\captionsetup[subfloat]{captionskip=-1pt}
\begin{center}
\vspace{-0.2in}
\hspace*{-0.55in}
\subfloat[]{\includegraphics[width=0.58\textwidth]{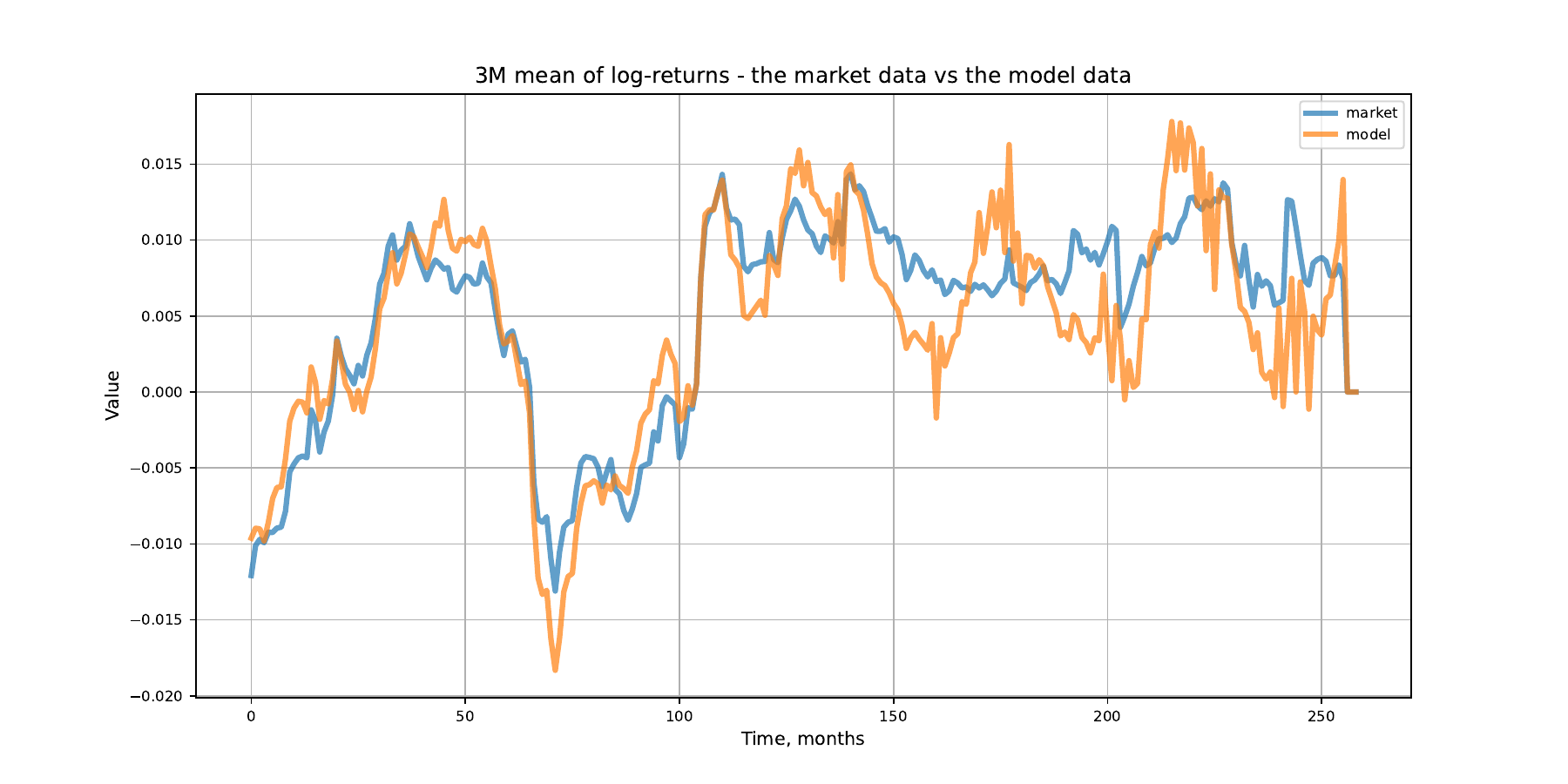}}
\hspace*{-0.4in}
\subfloat[]{\includegraphics[width=0.58\textwidth]{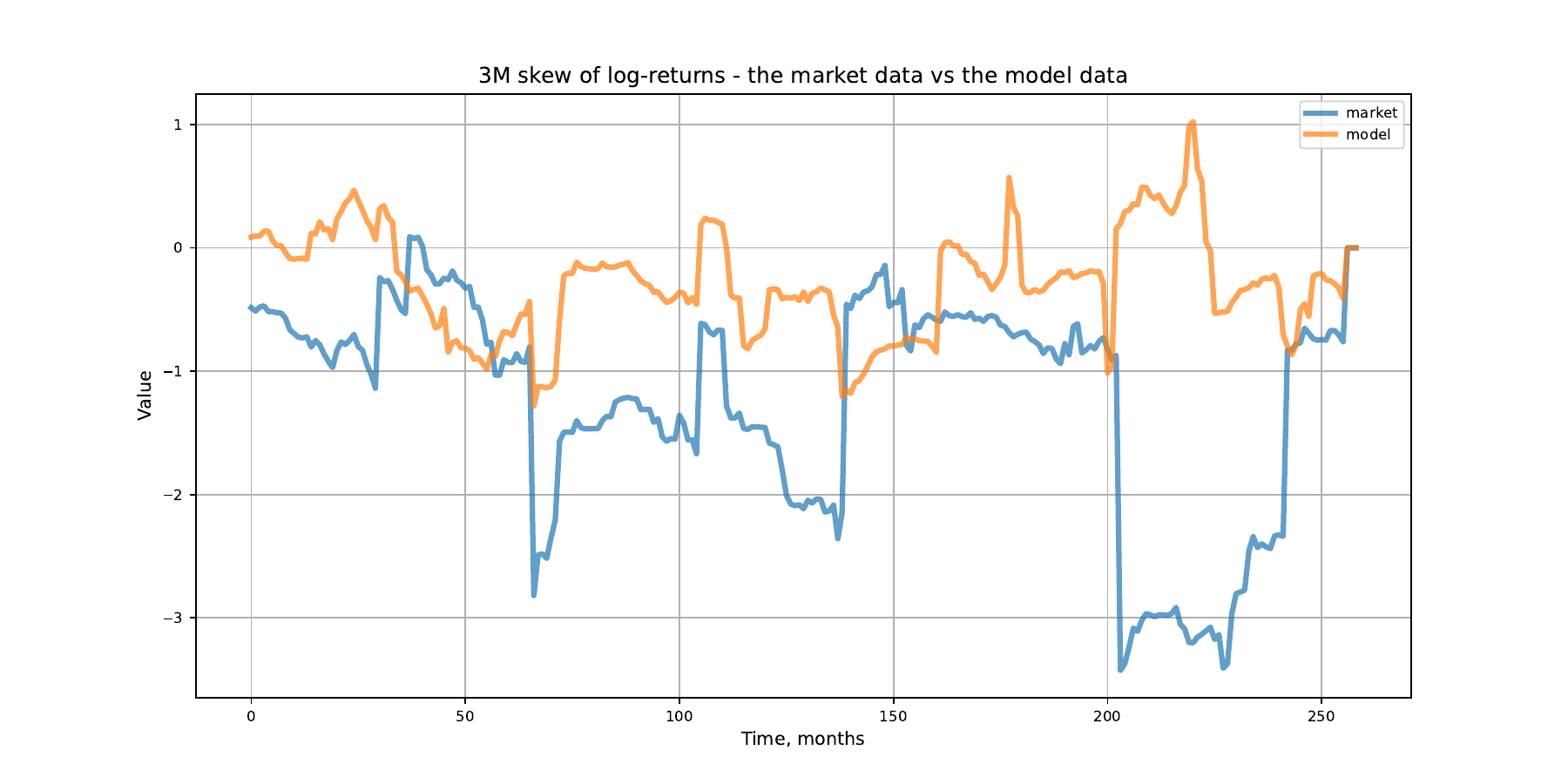}}

\hspace*{-0.55in}
\subfloat[]{\includegraphics[width=0.58\textwidth]{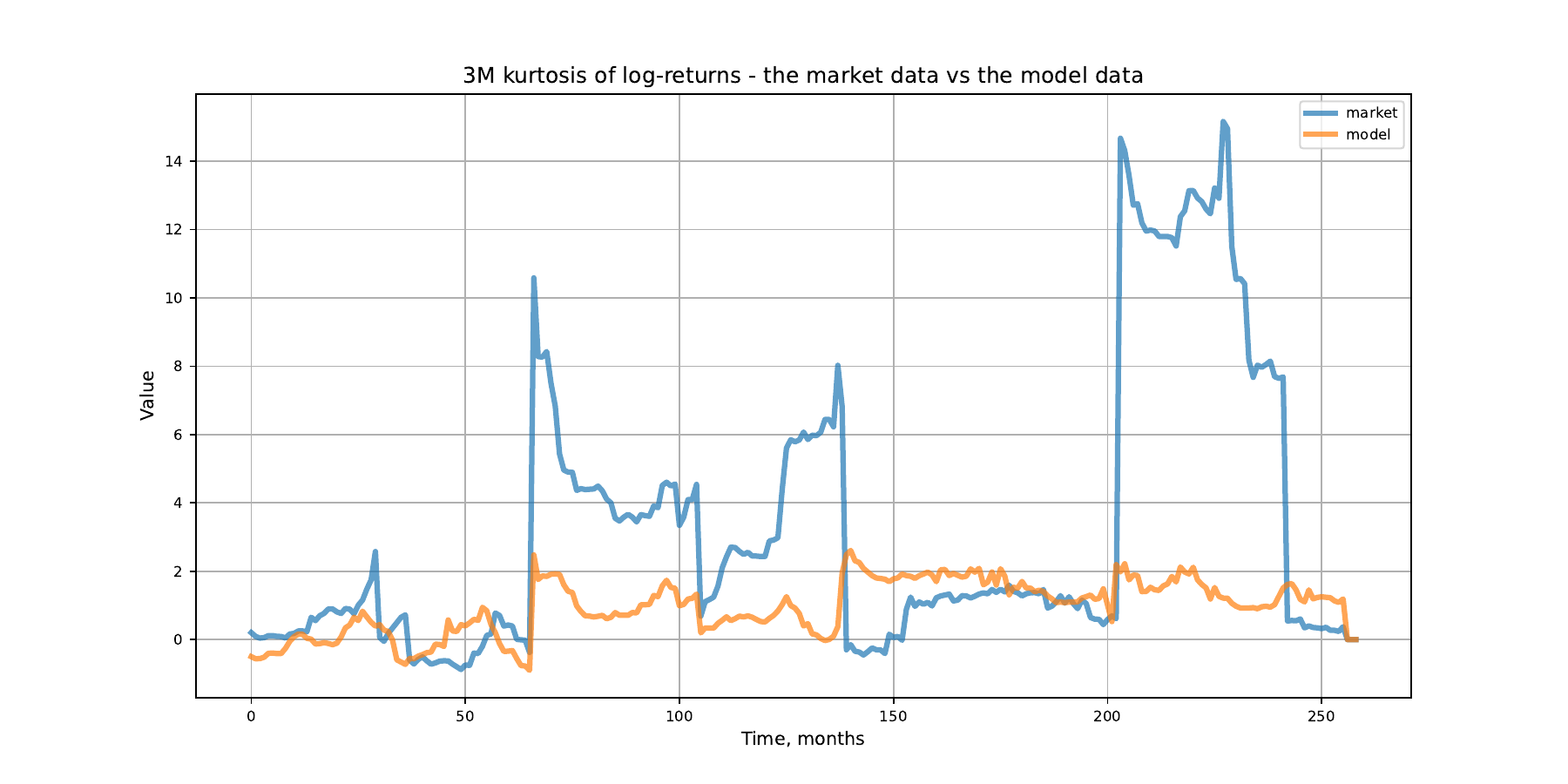}}
\hspace*{-0.4in}
\subfloat[]{\includegraphics[width=0.58\textwidth]{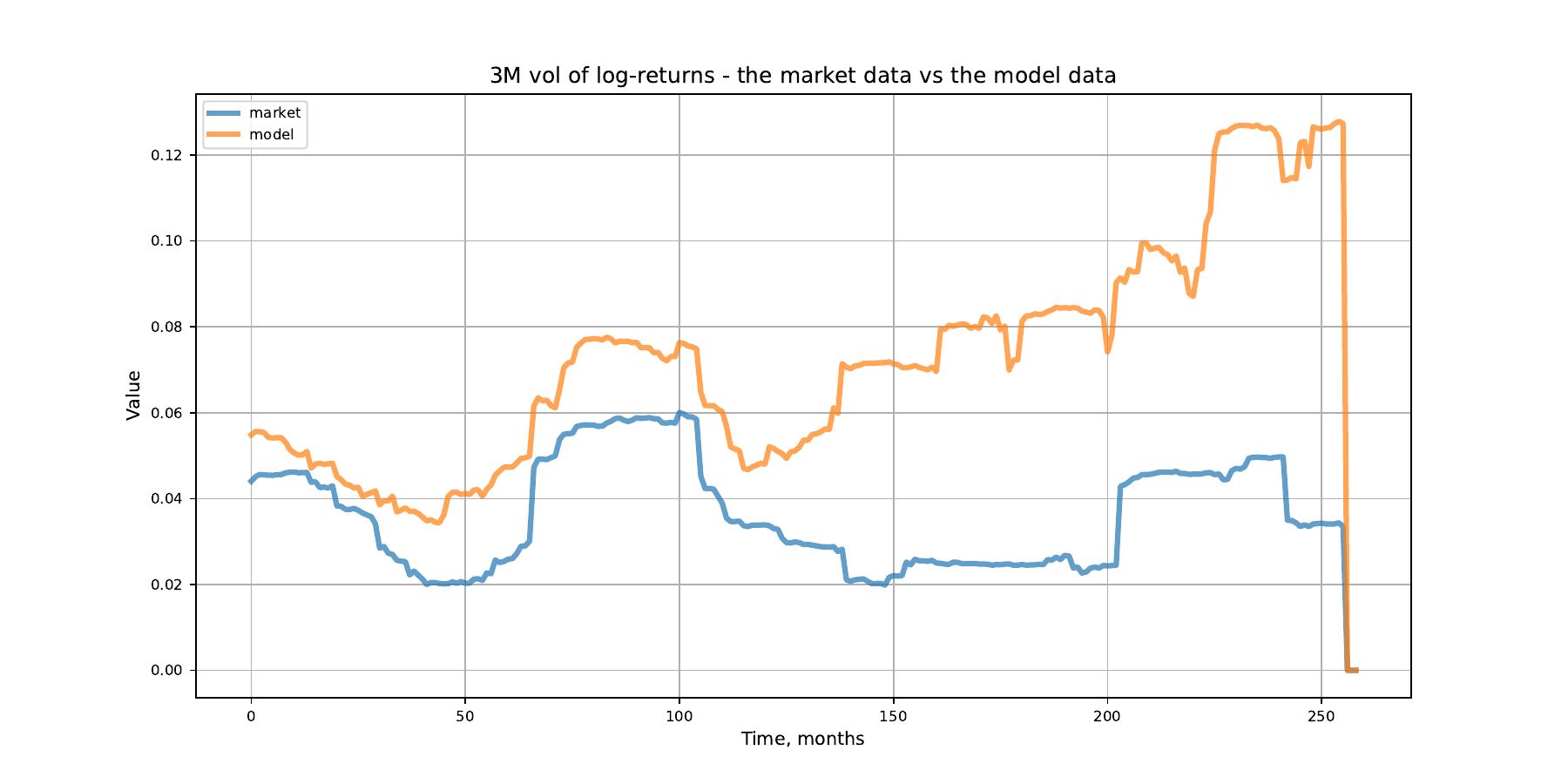}}
\end{center}
\caption{Three months moments of log-returns obtained by using the market and model data: a) mean, b) skew, c) kurtosis, and d) volatility.}
\label{rollingMoments}
\end{figure}

Therefore, we implement a simplified approach as follows. For each month $i$ from January 2003 to May 2024, we utilize historical monthly log-prices from two distinct periods: the preceding three years (36 points) and the subsequent three months (3 points). We then compute the first four moments at point $i+3$ using (i) market data and (ii) our model simulations. This approach effectively substitutes three months of daily historical data (approximately 60 points) with three years of monthly historical data (39 points). While this sample size is not large enough to generate robust statistics, it allows us to assess the model's ability to replicate market trends.

The results are presented in Fig.~\ref{rollingMoments}. Notably, the model's rolling mean closely tracks the market rolling mean, with divergence only in the most recent period. A similar pattern emerges for skewness, although there is a noticeable difference in levels that becomes particularly pronounced in recent periods. Regarding volatility and kurtosis, while the trends exhibit similar patterns, level differences become apparent in the right portion of the graph.

\subsubsection{The case $\theta_* = \hat{\theta}$}

\graphicspath{{./thetaHat/}}

As mentioned, this is the case of fast relaxation of the variable $\theta_t$. However, it differs from the previous case only by the values of the constraints (not the type!) imposed on the marketron potential during calibration. Accordingly, we expect the calibrated parameters to change, and the significance of these changes can be seen in Table~\ref{calibParamThetaHat}.
\begin{table}[!htb]
\begin{center}
\scalebox{0.87}{
\begin{tabular}{|l|r|r|r|r|r|r|r|r|r|}
\toprule
\rowcolor[rgb]{ .792,  .929,  .984}
{\bf parameter} & $\bm \sigma$ & $\bm \sigma_y$ & $\bm \sigma_z$ & $\bm \eta$ & $\bm k$ & $\bm \mu$ & $\bm g$ & ${\bm \hat{\theta}}$ & ${\bm \bar{y}}$ \\ \hline
lower bound & 0 & 0 & 0 & $-\sigma_{\max}^2/2$ & 0 & 0 & 0 & 0 & 0 \\ \hline
value       & 0.7743 & 0.8508 & 0.9524 & 0.0058 & 1.7684 & 1.4008& 0.3927 & 4.1076 & 0.7823  \\ \hline
upper bound & 3.0 & 3.0 & 3.0 & $1 - \sigma_{\min}^2/2$ & 5 & 3 & 1 & 10 & 1 \\ \hline
\rowcolor[rgb]{ .792,  .929,  .984}
{\bf parameter} & $\bm c$ & $\bm b_1$ & $\bm b_2$ & $\bm k_{1,x}$ & $\bm k_{2,x}$ & $\bm k_{3,x}$ & $\bm k_{1,y}$ & $\bm k_{2,y}$ & $\bm k_{3,y}$ \\ \hline
lower bound & 0 & -10 & -10 & -5 & 0 & -5 & -5 & 0 & -5 \\ \hline
value       & 3.9358 & 1.7983 & 2.4441 & 2.0011 & 1.4876 & -3.5391 & 3.5431 & 1.2359 & 0.1162 \\ \hline
upper bound & 5 & 10 & 10 & 5 & 5 & 5 & 5 & 5 & 5 \\
\bottomrule
\end{tabular}
}
\caption{Parameters of the model in \eqref{Marketron_3D} with the constraints in \eqref{finConstrRed} found by calibration to S\&P500 weekly returns from 2000 to Sept. 2024, together with the box constraints used in the calibration procedure.}
\label{calibParamThetaHat}
\end{center}
\end{table}

\graphicspath{{./thetaHat/}}
\begin{figure}[!htp]
\captionsetup[subfloat]{captionskip=-30pt}
\begin{center}
\scalebox{0.8} {
\hspace*{-0.9in}
\subfloat[]{\includegraphics[width=0.7\textwidth]{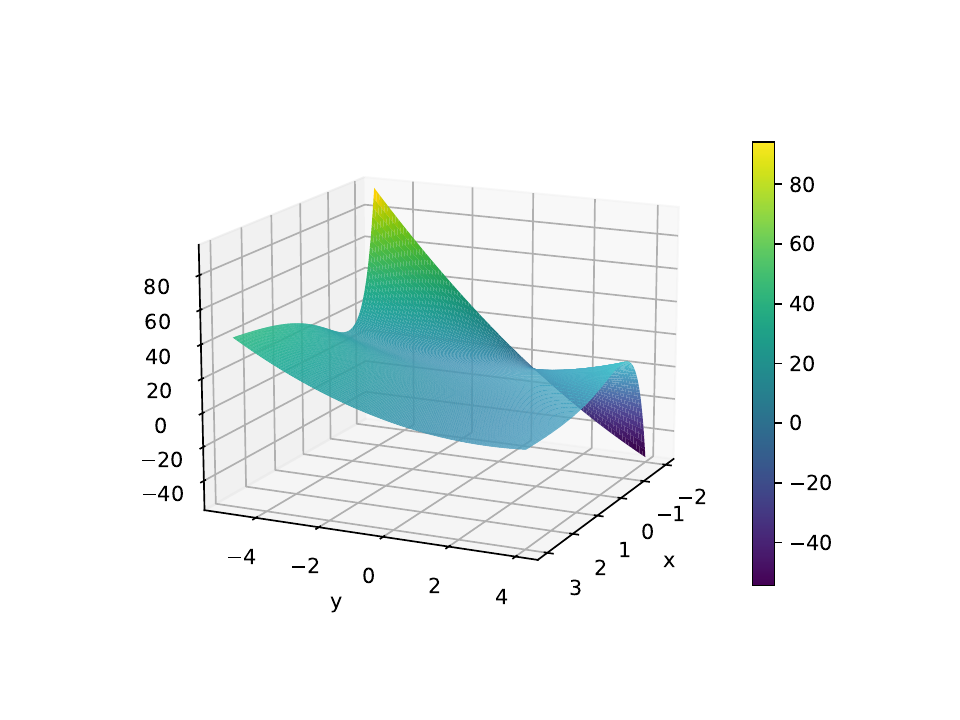}}
\hspace*{-0.4in}
\subfloat[]{\includegraphics[width=0.7\textwidth]{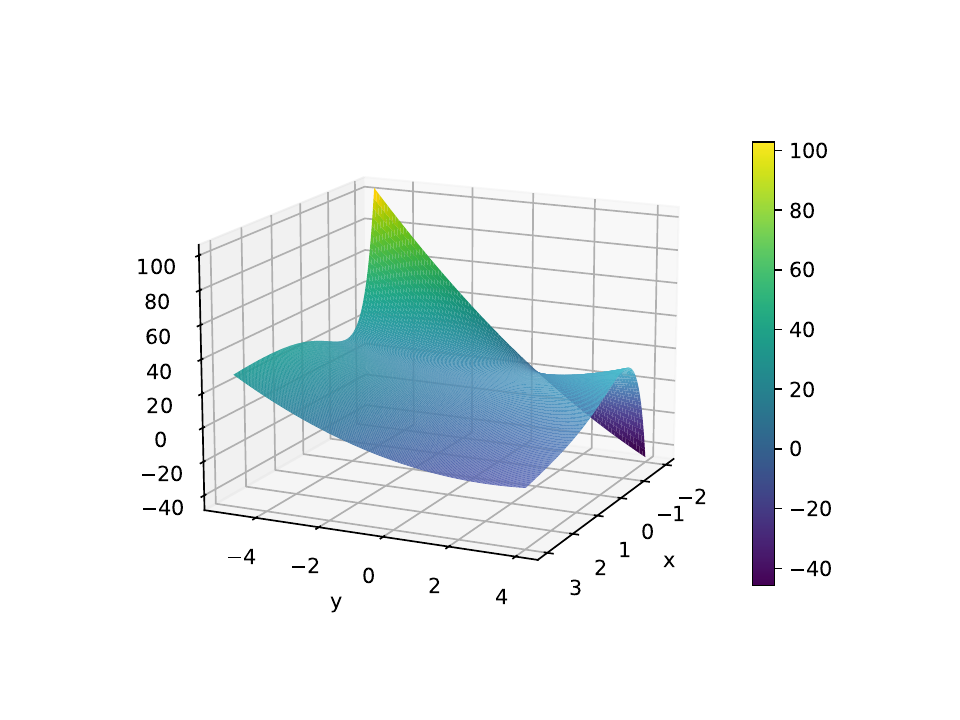}}
}
\end{center}
\caption{3D marketron potential $V(x,y)$ in \eqref{Lang_pot_general} computed with the model parameters found by calibration for a) $t=1$ year, b) $t=20$ years, $\theta_* = \hat{\theta}$.}
\label{pot3DcalibHat}
\end{figure}
\captionsetup[table]{skip=15pt}

Comparing the resulting model parameters with those in Table~\ref{calibParam} shows that while some parameters changed only slightly with the new $\theta_* = \hat{\theta}$, others exhibited significant differences. Our additional experiments revealed that these variations arose primarily from the objective function having multiple local minima rather than from the change in $\theta_*$ itself. These local minima, though distant from each other in parameter space, yield similar values of the objective function. As a result, the optimization procedure converges to different solutions depending on the random noise realization, while still maintaining good agreement with market data for the first four moments of the log-returns.
\begin{table}[!htb]
\centering
\scalebox{0.87}{
\begin{tabular}{|r|r|r|r|r|r|r|r|r|}
\toprule
\rowcolor[rgb]{ .949,  .808,  .937} \multicolumn{1}{|c|}{\textbf{hor, yrs}} & \multicolumn{1}{c|}{\textbf{mnMarket}} & \multicolumn{1}{c|}{\textbf{mnModel}} & \multicolumn{1}{c|}{\textbf{stMarket}} & \multicolumn{1}{c|}{\textbf{stModel}} & \multicolumn{1}{c|}{\textbf{skMarket}} & \multicolumn{1}{c|}{\textbf{skModel}} & \multicolumn{1}{c|}{\textbf{kuMarket}} & \multicolumn{1}{c|}{\textbf{kuModel}} \\ \hline
\rowcolor[rgb]{ .58, .863, .973} \textbf{2} & \cellcolor[rgb]{ .71, .902, .635} -0.2002 & \cellcolor[rgb]{ 1, 1, 1} -0.1847 & \cellcolor[rgb]{ .71, .902, .635} 0.1962 & \cellcolor[rgb]{ 1, 1, 1} 0.1958 & \cellcolor[rgb]{ .71, .902, .635} -0.1640 & \cellcolor[rgb]{ 1, 1, 1} -0.1686 & \cellcolor[rgb]{ .71, .902, .635} 0.1797 & \cellcolor[rgb]{ 1, 1, 1} 0.1842 \\ \hline
\rowcolor[rgb]{ .58, .863, .973} \textbf{5} & \cellcolor[rgb]{ .71, .902, .635} -0.0615 & \cellcolor[rgb]{ 1, 1, 1} -0.0597 & \cellcolor[rgb]{ .71, .902, .635} 0.1853 & \cellcolor[rgb]{ 1, 1, 1} 0.2117 & \cellcolor[rgb]{ .71, .902, .635} -0.1673 & \cellcolor[rgb]{ 1, 1, 1} -0.1468 & \cellcolor[rgb]{ .71, .902, .635} 0.1800 & \cellcolor[rgb]{ 1, 1, 1} 0.1838 \\ \hline
\rowcolor[rgb]{ .58, .863, .973} \textbf{10} & \cellcolor[rgb]{ .71, .902, .635} -0.0441 & \cellcolor[rgb]{ 1, 1, 1} -0.0408 & \cellcolor[rgb]{ .71, .902, .635} 0.2007 & \cellcolor[rgb]{ 1, 1, 1} 0.2272 & \cellcolor[rgb]{ .71, .902, .635} -0.3131 & \cellcolor[rgb]{ 1, 1, 1} -0.2935 & \cellcolor[rgb]{ .71, .902, .635} 0.3988 & \cellcolor[rgb]{ 1, 1, 1} 0.3951 \\ \hline
\rowcolor[rgb]{ .58, .863, .973} \textbf{15} & \cellcolor[rgb]{ .71, .902, .635} 0.0429 & \cellcolor[rgb]{ 1, 1, 1} 0.0441 & \cellcolor[rgb]{ .71, .902, .635} 0.1830 & \cellcolor[rgb]{ 1, 1, 1} 0.2009 & \cellcolor[rgb]{ .71, .902, .635} -0.3469 & \cellcolor[rgb]{ 1, 1, 1} -0.3477 & \cellcolor[rgb]{ .71, .902, .635} 0.4342 & \cellcolor[rgb]{ 1, 1, 1} 0.4672 \\ \hline
\rowcolor[rgb]{ .58, .863, .973} \textbf{20} & \cellcolor[rgb]{ .71, .902, .635} 0.0704 & \cellcolor[rgb]{ 1, 1, 1} 0.0663 & \cellcolor[rgb]{ .71, .902, .635} 0.1693 & \cellcolor[rgb]{ 1, 1, 1} 0.1822 & \cellcolor[rgb]{ .71, .902, .635} -0.3518 & \cellcolor[rgb]{ 1, 1, 1} -0.3665 & \cellcolor[rgb]{ .71, .902, .635} 0.4620 & \cellcolor[rgb]{ 1, 1, 1} 0.5290 \\ \hline
\rowcolor[rgb]{ .58, .863, .973} \textbf{24} & \cellcolor[rgb]{ .71, .902, .635} 0.0871 & \cellcolor[rgb]{ 1, 1, 1} 0.0810 & \cellcolor[rgb]{ .71, .902, .635} 0.1773 & \cellcolor[rgb]{ 1, 1, 1} 0.1794 & \cellcolor[rgb]{ .71, .902, .635} -0.3971 & \cellcolor[rgb]{ 1, 1, 1} -0.3272 & \cellcolor[rgb]{ .71, .902, .635} 0.4993 & \cellcolor[rgb]{ 1, 1, 1} 0.4931 \\
\bottomrule
\end{tabular}%
}
\caption{Annualized statistics of log-returns $x_t$ computed by using the model \eqref{Marketron_3D} with the model parameters found by calibration.}
\label{StatsXThetaHat}
\end{table}%

We note that differences in calibrated parameters only moderately affect the marketron potential shape, as shown in Fig.~\ref{pot3DcalibHat} (compare with Fig.~\ref{pot3Dcalib}). With either parameter set, the potential exhibits between 2 and 4 extrema.

Table~\ref{StatsXThetaHat} presents the Monte Carlo statistics using the newly calibrated parameters. Comparing these results with Table~\ref{Stats} indicates that setting $\theta_* = \theta_0$ produces almost the same alignment between market and model moments.

\section{Summary and outlook} \label{sect_Summary}

The paper introduces a new model of price formation in an inelastic market where dynamics are driven by both money flows and their price impact. This model extends previous work by one of the authors (see \cref{Introd} for detailed discussion). Unlike its predecessors, this model is multidimensional, incorporating three stochastic factors: the stock log-price $x_t$, the memory variable $y_t$, and the signal $z_t$.

The system dynamics are described by stochastic Langevin equations for $x_t$ and $y_t$, forming a multidimensional Langevin system.
The market price dynamics can thus be viewed as nonlinear diffusion of a particle (the marketron) in a two-dimensional space defined by the log-price $x$ and memory variable $y$. The system's potential function (the marketron potential) combines a Morse potential in $x_t$ with a harmonic potential in $y_t$. For certain choices of parameters, the marketron potential gives rise to instantons and metastable dynamics.

Section~\ref{neuron-sect} explores alternative interpretations of the marketron model. We demonstrate parallels between our model and certain neuroscience models of spiking neurons. Furthermore, we establish connections between our marketron model and physics-based models of active matter, highlighting the new interpretation of signals as the self-propelling components of the stock price dynamics.
This opens the way to new, physics-inspired approaches for finding optimal investment strategies for external investors as problems of control of active matter.

We explore various analytic approximations to our model in certain limits that reduce its effective dimensionality. Ultimately, we return to the complete problem and calibrate it using S\&P500 monthly log-prices from 2000 to Sept. 2024. The calibration involves solving a nonlinear optimization problem with nonlinear constraints, where the objective function minimizes a sum of the least squares error of the first four moments of the log-returns distribution across multiple time horizons (typically 5 to 9). We employ a particle filter to obtain time series of the hidden variables $y_t$ and $\theta_t$ (see \cref{calib}).

We observe that the model parameters found by calibration vary with different realizations of random noise, despite our use of a global search optimization algorithm. This variability can be attributed to multiple factors. First, the cross-entropy method implemented in the \verb|CEOpt| package is inherently stochastic. Second, the particle filter used to compute the objective function is also stochastic. Furthermore, our objective function appears to have numerous local minima that, while not proximate to each other in parameter space, yield nearly identical values of the objective function.

This problem can be addressed through several approaches. First, we can increase the number of particles in the particle filter while utilizing weekly or even daily market prices for comparison. Although this would significantly impact the calibrator's performance, implementing a parallel version of the particle filter could help mitigate these computational costs. Second, our problem involves many variables (eighteen), suggesting potential overfitting. One way to prevent or control for overfitting is to first calibrate the 1D approximate model of the marketron. Parameters common to both the 1D and 3D models could then be fixed, allowing the subsequent 3D calibration to focus only on the remaining variables. The third approach involves calibrating the model to SPY option prices rather than index data. This would require switching to a risk-neutral measure in equation \eqref{Marketron_3D}, deriving option price values (assuming a specific model for the market price of risk), and then calibrating these expressions to market option data. This approach will be explored in future work.

The technical challenges notwithstanding, the model demonstrates several strengths. First, it can be successfully calibrated to market data while producing the desired shape of the marketron potential. The revealed hidden states provide additional insights into the model's dynamics. The dynamics of transitions between metastable states (the Good, Bad and Ugly markets) resulting from our model resemble traditional regime-switching models popular in the financial literature, such as Hidden Markov models (HMMs), and share many features with these models, including volatility clustering and price-volatility correlations. Moreover, an HMM-type model can be viewed as an approximation of our model, with its parameters determined by the marketron potential parameters.


Furthermore, our model produces instantaneous defaults that can be observed during simulation, with default intensities determined by the model parameters. The calibrated marketron model generates an annualized default intensity of about $18$ bps, which reasonably aligns with market-implied default intensities typically ranging from $10$ to $50$ bps. This qualitative agreement is particularly noteworthy since we calibrated using only equity market data, without incorporating credit market information. While alternative parameterizations or calibration methods may yield different default intensities, we believe that the model's ability to capture default events without requiring a separate default framework is a significant advantage. This results in a more parsimonious approach where the natural relationship between equity and credit markets is inherently embedded into the model structure. A more precise alignment with credit market data could be achieved through joint calibration to both equity and credit indices — an extension we leave for future research.

\section*{Disclaimer}

Opinions expressed here are author's own, and do not represent views of their employers. A standard disclaimer applies.

\section*{Acknowledgments}

We thank Thierry Bollier, Dmitry Muravey and Thomas Schmeltzer for comments and helpful discussions.

\printbibliography[title={References}]

\appendix
\appendixpage
\appendix
\numberwithin{equation}{section}
\setcounter{equation}{0}

\section{Path integral for the Langevin dynamics} \label{sect_path_integral}

Consider again the Langevin equation \eqref{vector_Langevin} which we now re-write in the form common in physics
\beq \label{vector_Langevin_2}
\dot{\bm x} =  v {\bm n}_t - \nabla V({\bm x}) + \sigma {\bm \xi}_t, \qquad {\bm n}_t = \left(f(\theta_t), h(\theta_t) \right),
\eeq
\noindent where ${\bm \xi}_t$ is a two-dimensional white noise. Similarly, the OU equation \eqref{OU} with $K = 1$ can be written in a similar form
\beq \label{OU_theta}
\dot{\theta} = k( \bar{\theta} - \theta) + \sigma_z \boldepsilon_t,
\eeq
\noindent where $ \boldepsilon_t $ is another two-dimensional white noise. Probabilities of different paths generated by the Langevin equation \eqref{vector_Langevin_2} on the time interval $ [0, T] $ are determined by the joint Brownian path measure
\beq \label{Brownian_path_measure}
D \boldxi_t D \boldepsilon_t e^{- \frac{1}{2 \sigma^2} \int_{0}^{T} \boldxi_t^2 dt - \frac{1}{2 \sigma_z^2} \int_{0}^{T} \boldepsilon_t^2 dt },
\eeq
\noindent where $ D \boldxi_t D \boldepsilon_t  $ denotes the product of differentials $ d \boldxi_t, d\boldepsilon_t $ for all times $t \in [0,T]$.

Instead of integration over paths of the Brownian motions, we can change variables to the observable $ {\bf x}_t $ and $ \theta_t $.
This gives rise to a path integral expression for transition probabilities $ m_t({\bm x}_t, \theta_t | {\bm x}_0) $ to move
from the initial state ${\bm x}_0, \theta_0$ at time $t=0$ to the state ${\bm x}_t $ at time $t$
\beq \label{Langevin_path_integral_measure}
m_t({\bf x}_t | {\bf x}_0) = \int_{{\bf x}_0}^{{\bf x}_t}  D {\bf x} \int D \theta  e^{ -  \int_{0}^{T}  \mathcal{L}({\bf x}, \dot{\bf x}, {\bf \theta}, \dot{\bf \theta} dt },
\eeq
\noindent where $ \mathcal{L}({\bf x}, \dot{\bf x}, {\bf \theta}, \dot{\bf \theta}) $ is the Lagrangian
\beq \label{Langevin_Lagrangian}
\mathcal{L}({\bf x}, \dot{\bf x}, {\bf \theta}, \dot{\bf \theta}) = \frac{1}{2 \sigma^2}  \left| \left| \dot{\bf x} -  v {\bf n}_t + \nabla V({\bf x}) \right| \right|^2  + \frac{1}{2 \sigma_z^2} \left[ \dot{\theta} - k( \bar{\theta} - \theta) \right]^2
- \frac{\sigma^2}{2} \nabla^2 V + \lambda \left({\bf n}^2 - 1 \right).
\eeq
Here, the third term in the Lagrangian (proportional to $\sigma^2 $) is due to the Jacobian of the transformation from the Brownian measure \eqref{Brownian_path_measure} to the path integral measure, while the last term enforces the constraint $ {\bm n}^2 = 1 $ using the Lagrange multiplier $ \lambda $. Note that in \eqref{Langevin_path_integral_measure} we integrate over all trajectories of the hidden signal $ \theta_t $, while for the observed variable ${\bm x} $ the end points of all trajectories in the path integral are fixed.


In a weak noise limit $\sigma \to 0$, the escape from a metastable Bad market to another state via penetration through the barrier is modulated by the active signal $\theta_t$.
Such event can be modeled as the escape of a {\it self-propelled}, active particle from a metastable well (see \cref{sect_controlled_dynamics}). Such problems have recently drawn lots of interest in physics, see e.g., \cite{Woillez_2019, Chaki_2020, Gu_2020} and references therein.


When $\sigma \to 0$, the second and third terms in the Lagrangian \eqref{Langevin_Lagrangian} can be neglected. The most probable contributions to the path integral in this limit are given by trajectories that minimize the action functional in \eqref{Langevin_path_integral_measure}. These trajectories are obtained as the solution of the classical Euler-Lagrange equations of motion for fields ${\bf x}$ and ${\bf n}$.

Starting with the active signal field ${\bf n}$, in the limit $\sigma \to 0$ the contribution from the second term in \eqref{Langevin_Lagrangian} drops off. The Euler-Lagrange equation for ${\bf n}$ in this limit produces the following solution
\beq
\label{n_t}
{\bf n} = \frac{{\bf x} + \nabla V}{ \left| {\bf x} + \nabla V \right|}, \qquad \left| {\bf x} + \nabla V \right| = \sqrt{ \left| \left| {\bf x} + \nabla V \right| \right|^2}.
\eeq
Substituting this solution back into the Lagrangian \eqref{Langevin_Lagrangian}, we obtain a new effective Lagrangian for the weak noise limit, which is now a function of the ${\bf x}$-variable only
\beq
\label{eff_Lagrangian}
\mathcal{L}_\eff({\bf x}, \dot{\bf x}) = \frac{1}{2 \sigma^2}   \left( \left| \dot{\bf x} +  \nabla V \right| - v \right)^2.
\eeq

\section{Metastability in the D-limit}
\label{sect_metastability_in_1D}

In this appendix, we present an analytical approximation to computing
transition probabilities in the simplified 1D approximation to our full model that we introduced in Sect.~\ref{sect_The_Good_Bad_Ugly}. We develop this approximation in two steps. First, we approximate our 1D potential by a simpler 1D potential, and then we compute transition probabilities using this simplified potential instead of the actual potential arising in the 1D approximation of our model.

\subsection{Approximating $U_\eff(x)$ by a piecewise harmonic potential} \label{sect_approximate_potential}

\begin{figure}[!htbp]
\begin{center}
\includegraphics[width=0.6\linewidth]{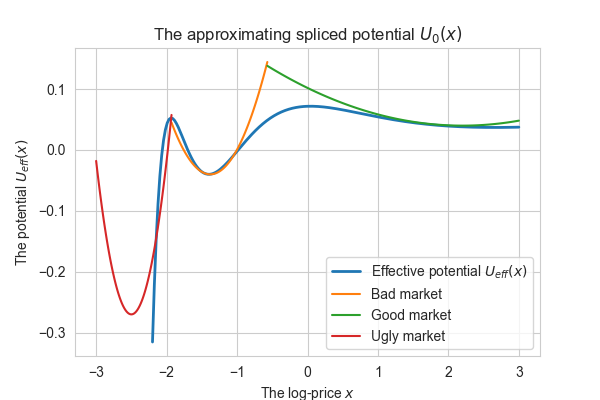}
\caption{Approximation of the marketron potential by a piecewise harmonic potential.}
\label{FigHarmon}
\end{center}
\end{figure}

In some cases one might wish to focus on the effective 1D dynamics in \eqref{Langevin_1D} with the effective 1D potential \eqref{pot_eff} obtained in the D-limit of the model. While \eqref{pot_eff} is represented by a simple analytical expression, it is not easily tractable. Therefore, we convert it to a more tractable formulation by replacing it with a piecewise harmonic potential (PHP) of the following form (see Fig.~\ref{FigHarmon})
\beq \label{U_PHP}
U_\eff(x) \to U_0(x) =
\begin{cases}
\dfrac{\omega_1^2}{2} \left(x - \bar{x}_1 \right)^2,        &-\infty \leq x \leq x_{L}, \\
\dfrac{\omega_2^2}{2} \left(x - \bar{x}_2 \right)^2 + u_2,  & \phantom{-} x_{L}  \leq x \leq x_{R}, \\
\dfrac{\omega_3^2}{2} \left(x - \bar{x}_3 \right)^2 + u_3,  & \phantom{-} x_{R}  \leq x \leq \infty.
\end{cases}
\eeq
Here $x_{L}$ and $x_{R} \geq x_{L}$ are two threshold values that separate different segments of the $x$-axis. The continuity of the potential at these points produces the following relations that can be used to fix parameters $u_2, u_3$ in terms of the remaining parameters of the PHP potential \eqref{U_PHP}
\begin{align} \label{continuity_U_PHP}
\frac{\omega_1^2}{2} \left(x_{L} - \bar{x}_1 \right)^2 &= \frac{\omega_2^2}{2} \left(x_{L} - \bar{x}_2 \right)^2 + u_2, \\
\frac{\omega_2^2}{2} \left(x_{R} - \bar{x}_2 \right)^2 + u_2 &= \frac{\omega_3^2}{2} \left(x_{R} - \bar{x}_3 \right)^2 + u_3. \nonumber
\end{align}
\graphicspath{{./Figs/}}

This approximation, however, produces a new effect that was not supposed to be a feature of the original model. Indeed, since the first derivative of the potential is only piecewise continuous, the drift of $x_t$ in \eqref{Langevin_1D} now contains weak singularities. As is known from (\cite{Portenko79,Lejay2006}, also see discussion in \cite{ItkinLiptonMuraveyA} and references therein), this gives rise to the process in \eqref{Langevin_1D} to be driven by a skewed Brownian Motion instead of a normal one. Thus, the log-prices under this approximation are skewed, while for the original marketron potential this is not obvious (though it can be, as we show in our numerical experiments).

\subsection{The Zwanzig approximation}

Per definition in \cref{sect_The_Good_Bad_Ugly}, the D-limit is characterized by an effective potential in \eqref{pot_eff}. We assume that the model parameters are such that the potential has a shape shown in Fig.~\ref{fig_Langevin_pot_1D_eff}, with a local minimum at $x = x_{\star}$ residing in the Bad market, and two local maxima $x_{L}, x_{R}$ with $x_{L} < x_{R}$, which describe the onset of the Ugly and Good markets, respectively. The local minimum at $x = x_{\star}$ is a metastable state; hence, for a given $z_t$ the market state $x_t$ can go over the top of either barrier to the left or to the right by an instanton transition. The effective potential \eqref{pot_eff} can be approximated by the PHP potential $U_0(x_t)$ in \eqref{U_PHP}.

Given a process $z_t$, according to \eqref{Langevin_1D} the marketron is moving in the full 1D potential
\beq \label{W_potential}
U(x_t) = U_0(x_t) + U_R(x_t), \qquad U_R(x_t) = -x_t z_t.
\eeq
This is a {\it random} potential combined from a deterministic function $U_0(x)$ of the random variable $x_t$ and another random component $U_R(x_t) = -x_t z_t$ which is controlled by the active signal $z_t$ evolving in accordance with the OU process \eqref{OU}.

Let $x_0$ be the initial position of marketron at time $t=0$ assumed to be located near the local minimum $x_{\star}$. Two different instanton transitions from this initial metastable state to one of the local maxima $x_{m} = \left\{ x_{R}, x_{L} \right\}$ can be considered. Accordingly, we can compute two different mean first passage times (MFPT) $\mathcal{T}_{BG}(x)$ and $\mathcal{T}_{BU}(x)$ for the escape through the right or left potential barrier, see e.g., \cite{Gardiner}
\beq \label{Tau}
\mathcal{T}_{BG}(x_0) = \frac{2}{\sigma^2} \int_{x_0}^{x_R} dx e^{ \frac{2 U(x)}{\sigma^2}} \int_{x_L}^{x} dy  e^{ - \frac{2 U(y)}{\sigma^2}}, \qquad
\mathcal{T}_{BU}(x_0) = \frac{2}{\sigma^2} \int_{x_L}^{x_0} dx e^{ \frac{2 U(x)}{\sigma^2}} \int_{x}^{x_R} dy  e^{ - \frac{2 U(y)}{\sigma^2}}.
\eeq
These expressions are valid for any potential $U(x)$ that is sufficiently well-behaved at $x \rightarrow \pm \infty$. In particular, they are valid if the potential has a single minimum. In this case, the values $x_{L}, x_{R}$ would not be the positions of the potential local maxima but rather threshold values of a log-return that signals a crisis or a stellar regime of the market. On the other hand, for scenarios where the Bad market is separated from the Good and Ugly markets by relatively tall barriers (as happens in the weak noise limit $\sigma \rightarrow 0$), Eqs.(\ref{Tau}) can be further simplified by replacing the inner integrals with a fixed-width integral over the interval $[x_L, x_R]$
\beq \label{MFPT1}
\mathcal{T}_{BG}(x_0) \simeq \frac{2}{\sigma^2} \int_{x_L}^{x_R} dy  e^{ - \frac{2 U(y)}{\sigma^2}} \int_{x_0}^{x_R} dx e^{ \frac{2 U(x)}{\sigma^2}}, \qquad
\mathcal{T}_{BU}(x_0) \simeq \frac{2}{\sigma^2} \int_{x_L}^{x_R} dy  e^{ - \frac{2 U(y)}{\sigma^2}} \int_{x_L}^{x_0} dx e^{ \frac{2 U(x)}{\sigma^2}}.
\eeq
As per \eqref{W_potential}, the exponential factors $e^{ - \frac{2 U(y)}{\sigma^2}}$ and $e^{\frac{2 U(x)}{\sigma^2}}$ in \eqref{MFPT1} depend on the active signals $z_t$. Following \cite{Chaki_2020}, we apply the method of \cite{Zwanzig_1988} to compute expectations of these exponents with respect to the active noise
\beq \label{active_noise_average}
e^{ - \frac{2 U(y)}{\sigma^2}} = e^{ - \frac{2 U_0(y)}{\sigma^2}} \left\langle e^{ \frac{2 y Z}{\sigma^2}} \right\rangle_{z}, \qquad
e^{  \frac{2 U(x)}{\sigma^2}} = e^{  \frac{2 U_0(x)}{\sigma^2}} \left\langle e^{ - \frac{2 x Z}{\sigma^2}} \right\rangle_{z},
\eeq
\noindent where $\langle \ldots \rangle_{z}$ stands for expectations with respect to a stationary distribution of the OU signal $z_t$. The Zwanzig method is applicable in our setting in the limit when the relaxation time of the signal $z_t$ is much smaller than the relaxation time for the log-returns that would be observed for a fixed value of the signal $z_t = z$.
As the stationary distribution of the OU process is Gaussian with zero mean and variance $\sigma_z^2/(2 \kappa)$, the expectations in \eqref{active_noise_average} are computed as follows
\beq \label{exp_exp_x}
e^{ - \frac{2 U_0(y)}{\sigma^2}} \left\langle e^{ \frac{2 y Z}{\sigma^2}} \right\rangle_{z} =
e^{ - \frac{2}{\sigma_2} \left( U_0(y) - \frac{\sigma_z^2}{2 \kappa \sigma^2} y^2 \right)},
\qquad
e^{  \frac{2 U_0(y)}{\sigma^2}} \left\langle e^{ - \frac{2 y Z}{\sigma^2}} \right\rangle_{z} =
e^{  \frac{2}{\sigma_2} \left( U_0(y) + \frac{\sigma_z^2}{2 \kappa \sigma^2} y^2 \right)}.
\eeq
Comparing these expressions with the PHP potential \eqref{U_PHP}, one can see that under the Zwanzig approximation, the net effect of the active noise $z_t$ reduces to renormalization of a piecewise quadratic potential \eqref{U_PHP}, which reads
\beq \label{renormalization_of_U_0}
\frac{2}{\sigma_2} \left( U_0(y) \pm \frac{\sigma_z^2}{2 \kappa \sigma^2} y^2 \right) = \Omega_{i}^{\pm} \left(y - \bar{y}_i^{\pm} \right)^2 + q_{i}^{\pm} \left( \bar{y}_{i}^{\pm} \right)^2 + \frac{2}{\sigma^2} u_i,
\eeq
\noindent where index $i$ takes values in $i \in [1,2,3] $ depending on the value of $y$ and according to segments defined in \eqref{U_PHP}. The parameters $ \bar{y}_i^{\pm}, \, \Omega_{i}^{\pm} $ and $ q_{i}^{\pm} $ are defined as follows
\beq \label{eff_params}
\bar{y}_i^{\pm} = \frac{\bar{x}_i}{1 - \frac{\sigma_z^2}{\kappa \sigma^2 \omega_i^2}}, \qquad
\Omega_{i}^{\pm} = \frac{\omega_i^2}{\sigma^2} \left( 1 - \frac{\sigma_z^2}{\kappa \sigma^2 \omega_i^2} \right), \qquad
q_{i}^{\pm} = \pm \frac{\sigma_z^2}{\kappa \sigma^2 \omega_i^2} \Omega_{i}^{\pm}.
\eeq
For a potential with two minima, the mean passage time can be calculated using a saddle point approximation. In this method, the first and second integrals in \eqref{Tau} are computed using quadratic expansions of $U(x)$ around a maximum $x_m$ and the local minimum $x^{\star}$, respectively. Such an approximation is justified when $\Delta U /h^2 \gg 1$, i.e., when the barrier is high, while the initial position $x_0$ resides near the local minimum $x^{\star}$. This gives rise to the celebrated Kramers escape rate formula for the escape intensity $\lambda = 1/\mathcal{T}$, see e.g., \cite{Gardiner,Hanggi_1986}
\beq \label{Kramers_rate_main}
\lambda = \frac{\sqrt{ U''(x^{\star} \left| U''(x_m) \right| }}{2 \pi} \exp \left[ - \frac{2 \Delta U }{h^2}\right].
\eeq
In particular, the Kramers escape rate $\lambda$ can be calculated as $\lambda = \Delta E/h^2$, where $\Delta E = E_{1}^{-} - E_0 = E_{1}^{-}$ is the energy splitting between the ground state and the first excited state. Within a path integral approach, the energy splitting $\Delta E$ can be obtained as a contribution to the path integral due to the instanton-saddle-point solutions of the dynamics obtained in a weak noise (quasi-classical) limit $h \rightarrow 0$ for the Langevin equation \eqref{Langevin_1D}, where the potential is inverted.

\section{Constraints on filtering inspired by the shape of marketron} \label{appConstr}

As mentioned in Sect.~\ref{sect_model_estimation}, we are looking for the marketron model parameters, which, on the one hand, are produced by calibration of the model to some market data, e.g., to the time series of some equity index, and, on the other hand, would preserve the shape of the potential depicted in Fig.~\ref{fig_Inverted_potential}. To achieve this, we want to impose some additional constraints to guarantee that the potential has three extrema. This appendix presents the derivation of such constraints.
In this derivation, we will use the full potential $V(x,y)$, meaning we add the term $v f(\theta)$ to $\eta$ to produce $\bar{\eta} = \eta + v f(\theta)$.

To proceed, we start by rewriting \eqref{Marketron_3D} in the form
\begin{align} \label{Marketron_3D_pot}
dx_t &=  -\fp{V}{x_t} dt + \sigma d W_t, \\
dy_t &= \left[ v h(\theta_t) - \fp{V}{y_t}  \right] dt + \sigma_y d \tilde{W}_t, \nonumber  \\
d \theta_t &= k( \hat{\theta} - \theta_t) dt + \sigma_{z} d Z_{t}, \nonumber
\end{align}
Here,
\begin{align} \label{potEqs}
- \fp{V}{x_t} &= \bar{\eta} - c(t) y_t V'_M(x_t), \\
- \fp{V}{y_t} &= \mu (\bar{y} - y_t) -  c(t) V_M(x_t), \nonumber
\end{align}
\noindent where $V_M(x)$ is defined in \eqref{U_pot_exact}. The \eqref{Marketron_3D_pot} is a 3D OU model with nonlinear drifts where all Brownian motions are independent.

Further, let us assume that the $y$-noise is small, so is $\sigma_y$. If in the first approximation we neglect the stochastic term in the equation for $y_t$, this equation becomes deterministic. Also, let us consider two cases where $h(\theta_t) = h(\hat{\theta})$ and $h(\theta_t) = h(\theta_0)$. In other words, for the $\theta_t$ variable here we consider two cases - of fast and slow relaxation. To cover both, let us denote the value in use as $\theta_*$, so $\theta_*$ could be either $\theta_0$ or $\hat{\theta}$.

With these assumptions, solving the second equation in \eqref{potEqs} with respect to $y(t)$\footnote{We now switch to the notation $y(t)$ instead of $y_t$ since in this case $y$ is a deterministic function.} and using the approximation in \eqref{U_pot_appr} yields
\begin{align} \label{y_sol}
y(t) &= I(t) - J(t) \left(e^{-x} - \frac{1}{2} g e^{-2 x} \right), \\
I(t) &= \left(\bar{y} + \frac{v h(\theta_*)}{\mu} \right) \left(1 - e^{- \mu t} \right) + y(0) e^{-\mu t}, \qquad
J(t) = c(t) \frac{1 - e^{- \mu t}}{\mu}. \nonumber
\end{align}
Substituting this solution into the first equation in \eqref{potEqs} yields
\begin{align} \label{potVx}
- \fp{V}{x} &= \bar{\eta} + c(t) e^{-x} \left( 1 - \frac{g e^{-x}}{1 + \varepsilon g e^{-x}} \right) \left[ I(t) - J(t) \left(e^{-x} - \frac{1}{2} g e^{-2 x} \right) \right].
\end{align}
Introducing a new variable $e^{-x} \mapsto z$, we obtain from \eqref{potVx}
\begin{align} \label{potVz}
z \fp{V}{z} &= \bar{\eta} + c(t) z \left(1 - \frac{g z}{1 + \varepsilon g z}\right) \left[ I(t) - J(t) z \left(1 - \frac{g}{2} z\right) \right].
\end{align}

Let us assume that the extrema of the potential lie in the finite area $z$, so we can neglect by the term $\varepsilon g \varepsilon$ in \eqref{potVx}. Then the RHS of \eqref{potVz} is a quartic polynomial of $z$, hence it might have 0, 2, 3 or 4 real roots. Definitely, we are interested in the latter three cases, which implies that the marketron potential $V(z)$ has two, three, or four extrema.

To proceed, let us represent this polynomial in the form
\begin{align} \label{cubic}
P(z) &= A(t) z^4 + B(t) z^3 + C(t) z^2 + D(t) z + E(t), \\
A(t) = -\frac{1}{2} g^2 c(t) J(t), \quad B(t) &= \frac{3}{2} g c(t) J(t), \quad C(t) = - c(t) \left[ g I(t) + J(t) \right], \quad D(t) = c(t) I(t) \quad E(t) = \bar{\eta}. \nonumber
\end{align}

Let us introduce the determinant of the quartic polynomial, \cite{Irving2004}
\begin{align} \label{det}
\Delta &= 256A^{3}E^{3} - 192A^{2}B D E^{2} - 128A^{2}C^{2} E^{2} + 144 A^{2}C D^{2}E - 27A^{2}D^{4} \\
&+ 144 A B^{2}C E^{2} - 6 A B^{2}D^{2}E - 80 A B C^{2}D E + 18ABCD^{3} + 16AC^{4}E \nonumber \\
&- 4AC^{3}D^{2} - 27B^{4}E^{2} + 18B^{3}CDE - 4B^{3}D^{3} - 4B^{2}C^{3}E + B^{2}C^{2}D^{2}, \nonumber
\end{align}
\noindent and three other polynomials
\begin{equation}
P = 8 A C - 3B^2, \quad D = 64 A^3 E - 16 A^2 C^2 + 16 AB^2 C - 16 A^2 BD - 3B^4, \quad \Delta_0 = C^2 - 3 B D + 12 A E.
\end{equation}
It is known that the possible cases for a quartic polynomial to have four real roots are as follows, \cite{Irving2004}
\begin{enumerate}
\item If $\Delta > 0$ and $P < 0, D < 0$, then all four roots are real and distinct.
\item If $\Delta = 0$ and $P < 0, D < 0, \Delta_0 \ne 0$, then there are a real double root and two real simple roots.
\item If $\Delta < 0$ there are two  distinct real roots and two complex conjugate roots.
\end{enumerate}
Computing the RHS of \eqref{det} with allowance for the definitions in \eqref{cubic} yields
\begin{align} \label{finConstr}
\Delta &= \frac{1}{16} c(t)^3 g^2 J(t) \Bigg\{
c(t)^3 g^2 I(t)^4 [J(t) - 2 g I(t) ] + 16 c(t) \bar{\eta}^2 g^2 J(t) \left[ -2 g^2 I(t)^2 - 2 g I(t) J(t)  + J(t)^2\right] \nonumber \\
&-4 c(t) \bar{\eta}  \Big[ c(t) g^4 I(t)^4 + 2 g^2 I(t)^2 J(t) \left[3 c(t) g I(t)  + 2 \right] + 8 J(t)^3 \left[ 3 c(t) g I(t)  + 2 \right] \\
&+ 2 g I(t) J(t)^2  \left[ 5 c(t) g I(t)  + 8 \right] + 16 c(t) J(t)^4 \Big] - 64 \bar{\eta}^3 g^4 J(t)^2
\Bigg\}, \nonumber \\
P &= c(t)^2 g^2 J(t) [g I(t)  - J(t)], \nonumber \\
D &= \frac{1}{8} c(t)^3 g^4 J(t)^2 \left[ 8 c(t) J(t) \left(g I(t)  + J(t) \right) + c(t)^2 \left( g I(t)  + 2 J(t) \right)^3 - 8 g^2 \bar{\eta} J(t)\right]. \nonumber
\end{align}

By definition in \eqref{u_t_general}, $c(t) > 0$. Therefore, in this case the necessary conditions for having four real roots are $\Delta > 0, P < 0, D < 0$ are reduced to the form
\begin{align} \label{finConstrRed1}
0 &> J(t) [g I(t) - J(t)], \quad \Delta > 0, \\
0 &> 8 c(t) J(t) \left[ g I(t) s + J(t) \right] + c(t)^2 \left[ g I(t) s + 2 J(t) \right]^3 - 8 g^2 \bar{\eta} J(t), \nonumber
\end{align}
And the other conditions are: $\Delta < 0$ for having two real roots, or $\Delta = 0, P < 0, D < 0, \Delta_0 \ne 0$ for having three real roots. These constraints are imposed on parameters of the marketron model when doing nonlinear filtering.

Recall that these conditions are obtained assuming small noise in the second equation in \eqref{Marketron_3D_pot}. Otherwise, the second equation in \eqref{Marketron_3D_pot}  can still be integrated to produce the same result as in \eqref{y_sol}. but where $y(0)$ is now replaced with
\begin{equation}
y(0) \mapsto y(0) - \sigma_y \int_0^t e^{k\mu} \epsilon_t dk.
\end{equation}
In this case the constraint in \eqref{finConstr} becomes stochastic, and it is not obvious how it can be used in the filtering method. Therefore, we impose an additional constraint that reads
\begin{equation}
y(0)  \gg \sigma_y \int_0^t e^{k\mu} \epsilon_t dk = \sigma_y \left( W_t e^{\mu t}  - \frac{1}{\mu} \int_0^t W_k e^{k\mu} dk \right).
\end{equation}
The RHS of this inequality is a martingale, therefore, on average, it is always true.

\end{document}